\documentclass[reprint, floatfix,nofootinbib,amsmath,amssymb,unsortedaddress,superscriptaddress]{revtex4-2}
\usepackage[utf8]{inputenc}
\usepackage{graphicx, cancel, booktabs, tabularx, amsmath, amsfonts, amssymb, bm, hyperref, placeins, float,mathrsfs}
\usepackage[normalem]{ulem}
\usepackage[table]{xcolor}

\newcolumntype{A}{>{\centering\arraybackslash}m{0.9 cm}}
\newcolumntype{B}{>{\centering\arraybackslash}m{0.5 cm}}
\newcolumntype{D}{>{\arraybackslash}m{6 cm}}

\newcolumntype{E}{>{\centering\arraybackslash}m{1.05in}}
\newcolumntype{L}{>{\centering\arraybackslash}m{0.25in}}
\newcolumntype{M}{>{\centering\arraybackslash}m{0.5in}}
\newcolumntype{N}{>{\centering\arraybackslash}m{1.35in}}

\newcommand{\ThetaD}{\Theta_\mathrm{D,\infty}}
\newcommand{\Vm}{V_\mathrm{m}}
\newcommand{\sigmaA}{\sigma^\mathrm{A}}
\newcommand{\omgG}{\omega_\mathrm{\Gamma,\,max}}

\newcommand{\kappaL}{\kappa_\mathrm{L}}
\newcommand{\kappaRT}{\kappa_L\left(\mathrm{300 K}\right)}

\newcommand{\kappaRTp}{\kappa^\mathrm{SISSO}\left(\mathrm{300 K}\right)}
\newcommand{\kappaRTpss}{\kappa^\mathrm{SISSO}_\mathrm{Slack}\left(\mathrm{300 K}\right)}

\usepackage{csvsimple}
\usepackage{longtable}
\usepackage{booktabs}

\begin{document}
\title{Accelerating Materials-Space Exploration for Thermal Insulators by Mapping Materials Properties via Artificial Intelligence}
\author{Thomas A. R. Purcell}
\email{purcell@fhi-berlin.mpg.de}
\affiliation{The NOMAD Laboratory at Fritz-Haber-Institut der Max-Planck-Gesellschaft and IRIS-Adlershof of the Humboldt-Universität zu Berlin, Faradayweg 4–6, D-14195 Berlin, Germany}

\author{Matthias Scheffler}
\affiliation{The NOMAD Laboratory at Fritz-Haber-Institut der Max-Planck-Gesellschaft and IRIS-Adlershof of the Humboldt-Universität zu Berlin, Faradayweg 4–6, D-14195 Berlin, Germany}
\affiliation{Physics Department and IRIS Adlershof Humboldt Universit\"at zu Berlin, Berlin, Germany.}

\author{Luca M. Ghiringhelli}
\email{ghiringhelli@fhi-berlin.mpg.de}
\affiliation{Physics Department and IRIS Adlershof Humboldt Universit\"at zu Berlin, Berlin, Germany.}
\affiliation{The NOMAD Laboratory at Fritz-Haber-Institut der Max-Planck-Gesellschaft and IRIS-Adlershof of the Humboldt-Universität zu Berlin, Faradayweg 4–6, D-14195 Berlin, Germany}

\author{Christian Carbogno}
\email{carbogno@fhi-berlin.mpg.de}
\affiliation{The NOMAD Laboratory at Fritz-Haber-Institut der Max-Planck-Gesellschaft and IRIS-Adlershof of the Humboldt-Universität zu Berlin, Faradayweg 4–6, D-14195 Berlin, Germany}

\begin{abstract}
Reliable artificial-intelligence models have the potential to accelerate the discovery of materials with optimal properties for various applications, including superconductivity, catalysis, and thermoelectricity.
Advancements in this field are often hindered by the scarcity and quality of available data and the significant effort required to acquire new data.
For such applications, reliable surrogate models that help guide materials space exploration using easily accessible materials properties are urgently needed.
Here, we present a general, data-driven framework that provides quantitative predictions as well as qualitative rules for steering data creation for all datasets via a combination of symbolic regression and sensitivity analysis.
We demonstrate the power of the framework by generating an accurate analytic model for the lattice thermal conductivity using only 75 experimentally measured values.
By extracting the most influential material properties from this model, we are then able to hierarchically screen 732 materials and find 80 ultra-insulating materials.
\end{abstract}
\date{\today}
\maketitle

\section{Introduction}
\label{sec:Intro}
\FloatBarrier

Artificial-intelligence~(AI) techniques have the potential to significantly accelerate the search for novel, functional materials, especially for applications where different physical mechanisms compete with each other non-linearly,~e.g.,~quantum materials~\cite{Stanev2021}, and where the cost of characterizing the materials makes a large-scale search intractable,~e.g.,~thermoelectrics~\cite{Miller2017a}.
Due to this inherent complexity, only limited amounts of data are currently available for such applications, which in turn severely limits the applicability and reliability of AI techniques~\cite{Gomes2019}.
Using thermal transport as an example, we propose a route to overcome this hurdle by presenting an AI framework that is applicable to scarce datasets and that provides heuristics able to steer further data creation into materials-space regions of interest.

Heat transport, as measured by the temperature-dependent thermal conductivity, $\kappa_\mathrm{L}$, is a ubiquitous property of materials and plays a vital role for numerous scientific and industrial applications including energy conversion~\cite{Zhang2013d}, catalysis~\cite{ChristianEnger2008}, thermal management~\cite{Wu2015}, and combustion~\cite{Pollock2016}.
Finding new crystalline materials with either an exceptionally low or high thermal conductivity is a prerequisite for improving these and other technologies or making them commercially viable at all.
Accordingly, finding new thermal insulators and understanding where in materials space to search for such compounds is an important open challenge in this field.
From a theory perspective, thermal transport depends on a complex interplay of different mechanisms, especially in thermal insulators, for which strongly anharmonic, higher-order effects can be at play~\cite{Knoop2020}.
Despite significant progress in the computational assessment of $\kappa_\mathrm{L}$ in solids~\cite{Broido2007ua,Carbogno2017}, these {\it ab initio} approaches are too costly for a large-scale exploration of material space.
For this reason, computational {\it high-throughput} approaches have so far covered only a small subset of materials~\cite{Carrete2014,Seko2015a,Xia2018}.
Experimentally, an even smaller number of materials have had their thermal conductivities measured, and less than 150 thermal insulators identified~\cite{Zhu2021, SpringerMaterials}.

Recently, increased research efforts have been devoted to leveraging AI frameworks to extend our knowledge in this field.
In particular, various regression techniques have been proven to successfully interpolate between the existing data and approximate $\kappa_\mathrm{L}$ using only simpler properties~\cite{Carrete2014,Zhang2018c, Chen2019b, Zhu2021}; however, using these techniques to extrapolate into new areas of materials space is a known challenge.
More importantly, the explainbility of these models is limited by their inherent complexity.
Physically motivated, semi-empirical models, e.g. the Slack model~\cite{Slack1979}, perform slightly better in this regard because they encapsulate information about the actuating mechanism.
Recent efforts have used AI to extend the capabilities of these models~\cite{Miller2017a, Yan2015a,Toberer2011,Zhang2018c} to increase their accuracy in estimating $\kappa_\mathrm{L}$.
However, the applicability of such models is still limited by the physical assumptions entering the original expressions~\cite{Miller2017a,Yan2015a}.
A general model that removes these assumptions and achieves the quantitative accuracy of AI approaches, while retaining the qualitative interpretability of analytical models, is however, still lacking.

In this work, we tackle this challenge by using a symbolic regression technique to quantitatively learn $\kappa_\mathrm{L}$, using easily calculated materials properties.
While symbolic regression methods are typically more expensive to train than other kernel based methods, such as Kernel Ridge Regression (KRR) and Gaussian Process Regression (GPR), their prediction errors are typically equivalent to other methods and their natural feature reduction and resulting analytical expressions make them a useful method for explainable AI, as further illustrated below~\cite{Wang2019}.
Furthermore, the added cost of training does not affect the evaluation time of the given models, meaning the extra time only has to be spent at the beginning.
The inherent uncertainty estimate in methods like GPR, allows for a prediction of where the resulting models are expected to perform worse; however, we also propose a method to get an ensemble uncertainty estimate for symbolic regression that can be applied more generally to these types of models.
We further exploit the feature reduction of SISSO and expand upon its interpretability by using a global sensitivity analysis method to distill out the key material properties that are most important for modelling $\kappa_\mathrm{L}$ and to find the conditions necessary for obtaining an ultra-low thermal conductivity.
From here, we use this analysis to learn the conditions needed to screen materials in each step of a  hierarchical, high-throughput workflow to discover new thermal insulators.
Using this workflow we can then establish qualitative design principles that lend themselves to general application across material space and use them to find 80 materials with an ultra-low $\kappa_\mathrm{L}$.

\section{Results}
\label{sec:results}

\subsection{Symbolic Regression Models for Thermal Conductivity}
\label{subsec:results_sisso}
For this study, we use the sure-independence screening and sparsifying operator (SISSO) method as implemented in the \textsc{SISSO++} code~\cite{Purcell2022}.
This method has been used to successfully describe multiple applications including the stability of materials~\cite{Schleder2020}, catalysis~\cite{Han2021}, and glass transition temperatures~\cite{Pilania2019}.
To find the best low-dimensional models for a specific target property, in our case the room temperature, lattice thermal conductivity, $\kappa_\mathrm{L}\left(\mathrm{300~K}\right)$, SISSO first builds an exhaustive set of analytical, non-linear functions, i.e. trillions of candidate descriptors, from a set of mathematical operators and primary features, the set of user-provided properties that will be used to model the target property.
Here we are focusing on room temperature data only because that is what is the most abundant in the literature and relevant for potential applications; however, some temperature dependence will be inherently included via the temperature dependence of our anharmonicity factor $\sigma^\mathrm{A}$.
For this application the primary features are both the structural and dynamical properties for seventy-five materials with experimentally measured $\kappa_\mathrm{L}\left(\mathrm{300~K}\right)$~\cite{Morelli,Slack1962,Martin1972,Takahashi1980,Turkes1980a,Gerlich1982,WILLIAMS1984,Valeri-Gil1993,Morelli1995,Hohl1999,Young1999,Li2003,Kawaharada2004,Villora2008,Toher2014b,Lu2015,Huang2016,Pantian2017,Chen2019b} (see Section~\ref{subsec:methods_sisso_primary_features} and~\ref{sec:si_experimental_kappa} for more details).
By using the experimentally measured values for $\kappa_\mathrm{L}$ we avoid the issues related to the inconsistent reliability of different approaches to calculating $\kappa_\mathrm{L}$ for different material classes~\cite{Xia2020a,Knoop2022}, and hopefully create a universal model for it.
For many of the materials of interest here the standard Boltzmann Transport approach will be unreliable~\cite{Xia2020a,Knoop2022}, but the fully anharmonic \textit{ab initio} Green Kubo approach is unnecessarily expensive to use for all materials~\cite{Knoop2022}.
Combining theoretical and experimental data in this way allows one to avoid both the cost or unreliability of calculating, $\kappa_\mathrm{L}$ and the challenges of experimentally synthesizing and characterizing candidate materials.
As long as all samples are consistent across each feature, AI and ML based models will adapt the computational features to the experimental target.

\begin{figure}
    \centering
    \includegraphics{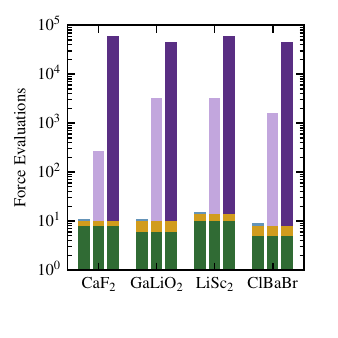}
    \caption{The motivation for the work is reducing the number of calculations needed to approximate the thermal conductivity of a material. a) The number of force evaluations needed to complete each step of a $\kappa_\mathrm{L}$ calculation for four representative materials: 1) Geometry relaxation (green first bar), 2) Harmonic model generation with Phonopy (yellow, second bar), 3) Evaluating $\kappa_\mathrm{L}$ via Phono3py (lavender third bar) or MD (purple fourth bar). The relaxation step typically acts on the primitive cells ($\sim$10 atoms) while all other are done on supercells with $\sim$200 or more atoms. The number of force evaluations for Phono3py assumes all displacements are needed to calculate the third order force constants for version 2.5.1 b) The proposed hierarchical workflow that can screen out materials before the final calculations.}
    \label{fig:motivation}
\end{figure}

Figure~\ref{fig:motivation}b illustrates the main goal of the work: to learn which primary features are important for modeling $\kappa_\mathrm{L}$ and what thresholds of those indicate where thermal insulators are present.
As a result the figure also represents the workflow used to calculate $\kappa_\mathrm{L}$ and generate the primary features for the model.
All of the data generated in this workflow will be calculated using \textit{ab initio} methods, with each step representing an increasing cost of calculation, as shown in Figure~\ref{fig:motivation}a.
The total cost of calculating these primary features is several orders of magnitude smaller than explicitly calculating $\kappa_\mathrm{L}$, either with the Boltzmann Transport Equation or aiGK.
While using only compositional and structural features would further reduce the cost of generating them, it comes at the expense of decreasing the reliability and explainability of the models.
A goal of this work is to learn the screening conditions needed to remove materials at each step of the workflow in Figure~\ref{fig:motivation}b and only perform the intensive $\kappa_\mathrm{L}$ calculations on the most promising materials.
Because of this, we feel that using the features generated from this workflow is the most logical set to use.
Importantly, as described in Section~\ref{subsec:methods_sisso_primary_features} we use a consistent and accurate formalism for calculating all features in this workflow, and therefore expect a quantitative agreement between these features and their experimental counterparts.
Even if this framework were restricted to explore only high-symmetry materials, the overall cost of the calculations in a supercell would be reduced by a factor of one hundred as shown by the non-green bars in Figure~\ref{fig:motivation}a.
In the more general case we would be able to screen closer to 1000 more materials using this procedure over the brute-force workflows of calculating $\kappa_\mathrm{L}$ for all materials.
With the learned conditions one could then create a prescreening procedure by learning models for each of the relevant structural or harmonic properties using only compositional inputs, and use those to estimate $\kappa_\mathrm{L}$~\cite{Foppa2022}; however, that is outside of the scope of this work.

\begin{figure*}
    \centering
    \includegraphics{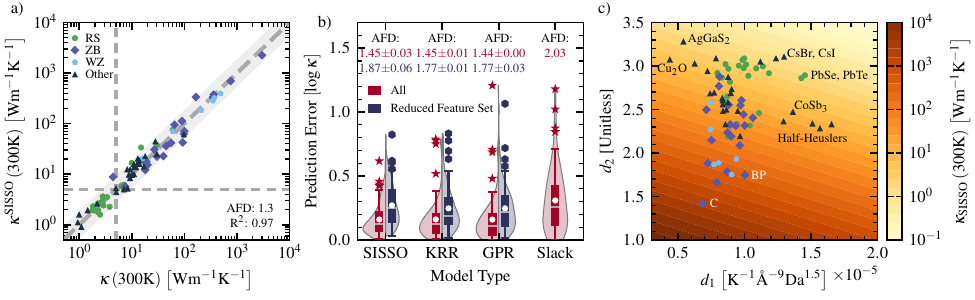}
    \caption{Error evaluation for the presented models. a) Comparison of the predicted $\kappa^\mathrm{SISSO}\left(\mathrm{300~K}\right)$ against the measured $\kappa_\mathrm{L}\left(\mathrm{300~K}\right)$ for the model trained against all data. The gray shaded region corresponds to the 95\% confidence interval. b) Violin plots of the mean prediction error of all samples for the SISSO, KRR, and GPR models using all features (red, left) and a reduced set including only $\sigma^\mathrm{A}$, $\Theta_\mathrm{D,\infty}$, and $V_\mathrm{m}$ (blue, right) and the Slack model. Gray lines are the median, white circles are the mean of the distributions, the boxes represent the quartiles, and the whiskers are the minimum and 95\% absolute error. For all calculations the parameterization depth and dimension are determined by cross-validation on each training set. The red stars and blue hexagons are the outliers for the box plots. c) A map of the two-dimensional SISSO model, where the features on the $x-$ and $y-$axes correspond to the two features selected by SISSO. The labeled points represent the convex-hull of the scatter plot and related points.}
    \label{fig:sisso_training}
\end{figure*}

In practice, we model the $\log\left(\kappa_\mathrm{L}\left(\mathrm{300~K}\right)\right)$ instead of $\kappa_\mathrm{L}\left(\mathrm{300~K}\right)$ itself to better handle the wide range of possible thermal conductivities.
The parity plot in Figure~\ref{fig:sisso_training}(a) illustrates the performance of the identified SISSO model when the entire dataset is used (see Section~\ref{subsec:methods_sisso} for more details).
The resulting expression is characterized by $d_1$ and $d_2$
\begin{equation}
\begin{split}
    \log \left(\kappa^\mathrm{SISSO}\left(\mathrm{300~K}\right)\right) &= a_0 + a_1 d_1 + a_2 d_2 \\
              d_1 &= \frac{ \left(m_\mathrm{avg}+200.3\, \mathrm{Da}\right)^2}{ \sqrt{ \mu }\left(V_\mathrm{m}+218.9 \,\text{\AA}^3\right)^3  \Theta_\mathrm{D,\infty} \sigma^\mathrm{A}} \\
              d_2 &= \sigma^\mathrm{A}\frac{ V_\mathrm{m} \rho}{ m_\mathrm{avg} } + \mathrm{e}^{\frac{-\omega_{\Gamma, \mathrm{max}}}{27.11\,\mathrm{THz}}} + \mathrm{e}^{ \sigma^\mathrm{A} }
\end{split}
\label{eqn:model}
\end{equation}
where $a_0=6.327$, $a_1=-8.219\times 10^4$, and $a_2=-1.704$ are constants found by least-square regression and all variables are defined in Table~\ref{tab:Primary Features}.
We find that this model has a training root-mean squared error (RMSE) of 0.14, with an $R^2$ of 0.98 for $\log \left(\kappa^\mathrm{SISSO}\left(\mathrm{300~K}\right)\right)$.
To better understand how these error terms translate to $\kappa_\mathrm{L}\left(\mathrm{300~K}\right)$, we also use the average factor difference (AFD)
\begin{subequations}
    \begin{align}
        \mathrm{AFD} &= 10^x \\
        x &= \frac{1}{n}\sum_{i}^n \left|\log\left(\kappa_\mathrm{L}\right) - \log\left(\kappa_\mathrm{L}^\mathrm{pred}\right)\right|,
    \end{align}
\end{subequations}
where $n$ is the number of training samples.
Here, we find an AFD of 1.30 that is on par if not smaller than models previously found by other methods (e.g. $1.36 \pm 0.03$ for a Gaussian Process Regression model~\cite{Chen2019b} and 1.48 for a semi-empirical Debye-Callaway Model~\cite{Miller2017a}).
However, differences in the training sets and cross-validation scheme prevent a fair comparison of these studies for the prediction error.
To see a complete representation of the training error for all models refer to ~\ref{sec:si_parity_comp}.

To get a better estimate of the prediction error, we use a nested cross-validation scheme further defined in Section~\ref{subsec:methods_error}.
As expected, the prediction error is slightly higher than the training error with an RMSE of $0.22\pm0.02$ and an AFD of $1.45\pm0.03$.
As shown in Fig.~\ref{fig:sisso_training}(b), these errors are comparable to those of a KRR and GPR model trained on the same data, following the procedures listed in Sections~\ref{subsec:methods_krr} and~\ref{subsec:methods_gpr}, respectively.
We chose to retrain the models using the same dataset and cross-validation splits in order to single out the effect of the methodology itself, and not changes in the data set and splits.
These results show that the performance of SISSO and more traditional regression methods are similar, but the advantage of the symbolic regression models is that only seven of the primary features are selected.
Another advantage of the nested cross-validation scheme is that it creates an ensemble of independent models, which can also be used to approximate the uncertainty of the predictions.
These results substantiates that our symbolic regression approach performs as well as interpolative methods and outperform the Slack model, which was originally developed for elemental cubic solids~\cite{Slack1979}.
Interestingly, offering the features of the Slack model to SISSO does not improve the results, and even some primary features previously thought to be decisive, e.g., the Grüneisen parameter, $\gamma$. are not even selected by SISSO (see~\ref{sec:si_slack_training}).

A key advantage of using symbolic regression techniques over interpolative methods such as KRR and GPR is that the resulting models not only yield reliable quantitative predictions, but also allows for a qualitative inspection of the underlying mechanisms.
To get a better understanding of how the thermal conductivity changes across materials space we map the model in Figure~\ref{fig:sisso_training}c.
From this map we can see that the thermal conductivity of a material is mostly controlled by $d_2$ with $d_1$ providing only a minor correction.
While these observed trends are already helpful, the complex non-linearities in both $d_1$ and $d_2$ impedes the generation of qualitative design rules.
Furthermore, some primary features such as $V_\mathrm{m}$ and $\sigma^\mathrm{A}$ enter both $d_1$ and~$d_2$,  with contrasting trends,~e.g.,~$\sigma^\mathrm{A}$ lowers $d_1$ but increases $d_2$.
To accelerate the exploration of materials space, one must first be able to disentangle the contradicting contributions of the involved primary features.

\subsection{Extracting Physical Understanding by identifying the Most Physically Relevant Features via Sensitivity Analysis}
\label{subsec:results_SA}
The difficulties in interpreting the ``plain'' SISSO descriptors described above can be overcome by performing a sensitivity analysis or a feature importance study to identify the most relevant primary features that build $d_1$ and~$d_2$.
For this purpose, we employ both the Sobol indices, i.e., the main effect index $S_i$ and the total effect index~$S_i^\mathrm{T}$~\cite{Sobol1993}, and the Shapley Additive Explanations (SHAP)~\cite{NIPS2017_7062} metric for the model predictions.
To calculate the Sobol indices we use an algorithm that includes correlative effects first described by Kucherenko {\em et al}.~\cite{Kucherenko2012}, and later implemented in \textsc{UQLab}~\cite{Marelli2014,Wiederkehr2018}.
The main advantage of this approach is its ability to include correlative effects between the inputs, which if ignored can largely bias or even falsify the sensitivity analysis results~\cite{Razavi2021}.
Qualitatively, $S_i$ quantifies how much the variance of $\log\left(\kappa_\mathrm{L}\left(\mathrm{300~K}\right)\right)$ correlates with the variance of a primary feature, $\hat{x}_i$,
and $S_i^\mathrm{T}$ quantifies how much the variance of $\log\left(\kappa_\mathrm{L}\left(\mathrm{300~K}\right)\right)$ correlates with $\hat{x}_i$ including all interactions between $\hat{x}_i$ and the other primary features.
For example, Sobol indices of 0.0 indicate that $\log\left(\kappa_\mathrm{L}\left(\mathrm{300~K}\right)\right)$ is fully independent of $\hat{x}_i$,  whereas a value of 1.0 indicates that $\log\left(\kappa_\mathrm{L}\left(\mathrm{300~K}\right)\right)$ can be completely represented by changes in $\hat{x}_i$~\cite{Wiederkehr2018}.
Moreover, $S^\mathrm{T}_i < S_i$ implies that correlative effects are significant, with an $S^\mathrm{T}_i=0$ indicating that a primary feature is perfectly correlated to the other inputs~\cite{Wiederkehr2018}.

The SHAP values constitute a local measure of how each feature influences a given prediction in the data set.
This metric is based on the Shapley values used in game theory for assigning payouts to players in a game based on their contribution towards the total reward~\cite{NIPS2017_7062}.
In the context of machine learning models each input to the model represents the players and the difference between individual predictions from the global mean prediction of a dataset represents the payouts~\cite{Aas2021}.
The SHAP values then perfectly distribute the difference from the mean prediction to each feature for each sample, with negative values indicating that the feature is responsible for reducing the prediction from the mean and a positive value is responsible for increasing it.~\cite{Aas2021}.
A similar metric is the Local Interpretable Model-agnostic Explanations (LIME) values~\cite{lime}.
LIME first defines a local neighborhood for each data point, and then uses a similar algorithm to SHAP to compare each prediction against their corresponding local area.
Because of the computational complexity of calculating SHAP values makes their exact calculation intractable with a large number of features, these values can be approximated by the Kernel SHAP method~\cite{NIPS2017_7062}.
Originally the Kernel SHAP method assumed feature independence~\cite{NIPS2017_7062}, but was recently advanced to include feature dependence via sampling over a multivariate distribution represented by a set of marginal distributions and a Gaussian Copula~\cite{Aas2021}.
However, there are some cases for small data sets with highly correlated features where the SHAP values are qualitatively different from the true Shapley values~\cite{Roder2021}.

\begin{figure*}
    \centering
    \includegraphics{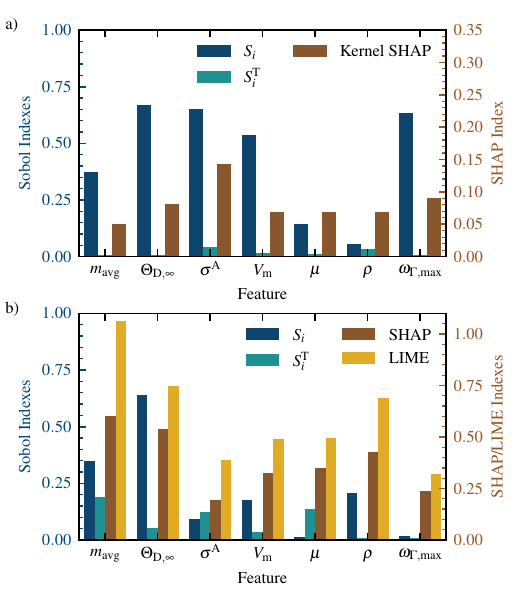}
    \caption{The feature importance metrics for the models. $S_i$ (first bar, dark blue), $S_i^T$ (second bar, light blue), mean absolute SHAP index (third bar, brown), and LIME index (fourth bar, yellow) for each feature in the model by treating the inputs as a) dependent feature and b) independent features. The Sobol indices are plotted on the left y-axis and the SHAP and LIME indexes are plotted on the right y-axis}
    \label{fig:sensitivity_analysis}
\end{figure*}

Figure~\ref{fig:sensitivity_analysis} compares the different sensitivity metrics including and excluding feature dependence.
To get the global values of the SHAP and LIME indexes we take the mean absolute value for each feature across all 75 materials, but other metrics have been proposed in the literature and it is not clear which one is best~\cite{Lee2022,Ittner2021,10.1145/3307339.3343255}.
However the local information contained in metrics such as SHAP and LIME is an advantage they have over global metrics such as the Sobol indexes as it allows for the identification of regions in the material space that do not follow the global trends.
Comparing the plots in Figure~\ref{fig:sensitivity_analysis}a and b illustrates the importance of not treating the input primary features as independent, as all four sensitivity analysis metrics are qualitatively wrong under that assumption.
This is likely a result of sampling over physically unreachable parts of the feature space, e.g. a areas with a high density, low mass, and high molar volume, and suggests that caution should be used when applying these techniques to highly correlated datasets.
The impact of this is demonstrated in Supplementary Figure~\ref{fig:si_sensitivity_analysis}, where we explicitly simplify the model to remove some of the dependencies.
All three indexes that include correlative effects show that $\sigma^\mathrm{A}$, $V_\mathrm{m}$, $\Theta_\mathrm{D,\infty}$, and $\omega_\mathrm{\Gamma, max}$ predominately control the variance of $\kappa^\mathrm{SISSO}\left(\mathrm{300~K}\right)$.
The main difference between $S_i$ and the kernel SHAP metrics is the relative importance of $\Theta_\mathrm{D,\infty}$ and $\omega_\mathrm{\Gamma,max}$ when compared against $V_\mathrm{m}$ and $\sigma^\mathrm{A}$.
The difference between these results could be from the the Sobol indexes globally sampling the region of $\Theta_\mathrm{D,\infty} > 1300$ K instead of relying on the two materials in that regime or $S_i$ over-estimating its importance because the higher correlation between $\Theta_\mathrm{D,\infty}$ and the other inputs.
In fact, the low values of $S^\mathrm{T}_i$ also imply that there are significant correlative effects in place between these inputs, and no single feature can be singled out as primarily responsible for changes in $\kappa^\mathrm{SISSO}\left(\mathrm{300~K}\right)$.
For instance, the similarity between the importance of $\omega_\mathrm{\Gamma, max}$ and $\Theta_\mathrm{D,\infty}$ is because they are strongly correlated to each other, only one of them needs to be considered (see the Supplementary Figure~\ref{fig:si_all_maps}).
The importance of these features is further substantiated in Figure~\ref{fig:sisso_training}b, where we compare the performance of the models calculated using the full dataset and one that only includes $\sigma^\mathrm{A}$, $V_\mathrm{m}$, and $\Theta_\mathrm{D,\infty}$.
For all tested models, we see only a slight deterioration in performance with a predictive AFD of 1.87, 1.77, and 1.77 for the SISSO, KRR, and GPR models, respectively, compared to 1.45 for the models trained with all features.
This result highlights  that the trends and the underlying mechanisms describing the dependence of $\kappa_\mathrm{L}\left(\mathrm{300~K}\right)$ in materials space are fully captured by those features alone.

\begin{figure*}[b!]
    \centering
    \includegraphics{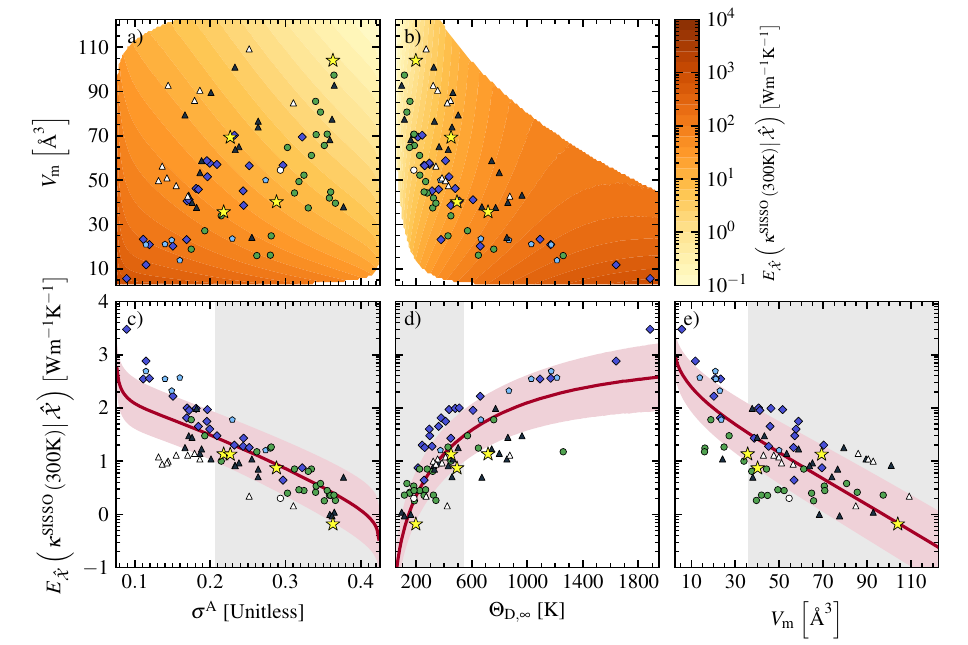}
    \caption{The expected value of $\kappa^\mathrm{SISSO}\left(\mathrm{300~K}\right)$ relative to select primary features. The expected value of $\kappa^\mathrm{SISSO}\left(\mathrm{300~K}\right)$, $E_{\hat{\mathcal{X}}}\left(\left.\kappa^\mathrm{SISSO}\left(\mathrm{300~K}\right)\right| \hat{\mathcal{X}} \right)$, where $\hat{\mathcal{X}}$ is a) $\left\{\sigma^\mathrm{A}, V_\mathrm{m}\right\}$, b) $\left\{\Theta_\mathrm{D,\infty}, V_\mathrm{m}\right\}$, c) $\left\{\sigma^\mathrm{A}\right\}$, d) $\left\{\Theta_\mathrm{D,\infty}\right\}$, and e) $\left\{V_\mathrm{m}\right\}$. $E_{\hat{\mathcal{X}}}\left(\left.\kappa^\mathrm{SISSO}\left(\mathrm{300~K}\right)\right| \hat{\mathcal{X}} \right)$ is calculated by sampling over the multivariate distributions used for the sensitivity analysis, and binning the input data until there are at least $10\,000$ samples in each bin. The red line in c-e corresponds to $E_{\hat{\mathcal{X}}}\left(\left.\kappa^\mathrm{SISSO}\left(\mathrm{300~K}\right)\right| \hat{\mathcal{X}} \right)$ and the pink shaded region is one standard deviation on either side of the line. The gray shaded regions represent where a thermal conductivity of 10 Wm$^{-1}$K$^{-1}$ or lower is within one standard deviation of the expected value. On all maps all materials in the training set are displayed. The green circles correspond to rock-salts, the blue diamonds are zincblende, the light blue pentagons are wurtzites, and black triangles are all other materials. All points with a $\kappa_\mathrm{L}\left(\mathrm{300~K}\right)$ less than one standard deviation below the expected value based on $\sigma^\mathrm{A}$ are highlighted in white. The points in c-e correspond to the actual values of $\kappa_\mathrm{L}\left(\mathrm{300~K}\right)$ for each material. Additionally we include four materials outside of the training set (yellow stars) whose thermal conductivities we calculate using \textit{ab initio} molecular dynamics.}
    \label{fig:maps}
\end{figure*}

Even more importantly, our model captures the interplay between these features across materials, as demonstrated in the maps in Figure~\ref{fig:maps}.
These maps showcase the strong correlation between $\kappa^\mathrm{SISSO}\left(\mathrm{300~K}\right)$ and $\sigma^\mathrm{A}$, $V_\mathrm{m}$, and $\Theta_\mathrm{D,\infty}$, and that materials with high anharmonicity, low-energy vibrational modes, and a large molar volume will be good thermal insulators.
Figure~\ref{fig:maps} shows the expected value of $\kappa^\mathrm{SISSO}\left(\mathrm{300~K}\right)$, $E_{\hat{\mathcal{X}}}\left(\left.\kappa^\mathrm{SISSO}\left(\mathrm{300~K}\right)\right| \hat{\mathcal{X}} \right)$, for different sets of input features, $\hat{\mathcal{X}}$, shown on the axes of each plot.
We then overlay the maps with the actual values of each input for all materials in the training set to evaluate the trends across different groups of materials.
Figure~\ref{fig:maps}c confirms that $\sigma^\mathrm{A}$ is already a good indicator for finding thermal insulators, with most of the materials having $\kappa_\mathrm{L}\left(\mathrm{300~K}\right)$ within one standard deviation of the expected value.
For the more harmonic materials with $\sigma^\mathrm{A}<0.2$, the vanishing degree of anharmonicity is, alone, not always sufficient for quantitative predictions.
In this limit, a combination of $\sigma^\mathrm{A}$ and $V_\mathrm{m}$ can produce correct predictions for the otherwise underestimated white triangles with a $\sigma^\mathrm{A}<0.2$, as seen in Figure~\ref{fig:maps}a.
In order to fully describe the low thermal conductivity of the remaining highlighted materials both $\Theta_\mathrm{D,\infty}$ and $V_\mathrm{m}$ are needed as can be seen in Figure~\ref{fig:maps}a, b, d and e.
Generally, this reflects that the \textbf{three} properties $\sigma^\mathrm{A}$, $\Theta_\mathrm{D,\infty}$, and $V_\mathrm{m}$ are the target properties to optimize to obtain ultra-low thermal conductivities.

These results can also be rationalized within our current understanding of thermal transport and showcase which physical mechanisms determine $\kappa_\mathrm{L}$ in material space.
Qualitatively, it is well known that good thermal conductors typically exhibit a high degree of symmetry with a smaller number of atoms,~e.g. diamond and silicon, whereas thermal insulators, e.g.,~glass-like materials, are often characterized by an absence of crystal symmetries and larger primitive cells.
In our case, this trend is quantitatively captured via $V_\mathrm{m}$, which reflects that larger unit cells have smaller thermal conductivities.
Furthermore, it is well known that phonon group velocities determine how fast energy is transported through the crystal in the harmonic picture~\cite{peierls1955quantum}, and that it is limited by scattering events arising due to anharmonicity.
In our model, these processes are captured by $\Theta_\mathrm{D,\infty}$, which describes the degree of dispersion in the phonon band structure, and the anharmonicity measure, $\sigma^\mathrm{A}$ respectively.
In this context, it is important to note that, in spite of the fact that these qualitative mechanisms were long known, there had hitherto been no agreement on which material property would quantitatively capture these mechanisms best across material space.
For instance, both the $\gamma$, the lattice thermal expansion coefficient, and now $\sigma^\mathrm{A}$, have been used to describe the anharmonicity of a material.
However, when both $\gamma$ and $\sigma^\mathrm{A}$ are included as primary features, only $\sigma^\mathrm{A}$ is chosen (see~\ref{sec:si_slack_training} for more details).
This result indicates that the $\sigma^\mathrm{A}$ measure is the more sensitive choice for modeling the strength of anharmonic effects.
While $\gamma$ also depends on anharmonic effects, they are also influenced by the bulk modulus, the density, and the specific heat of a material.

\subsection{Validating the Predictions with \textit{ab initio} Green-Kubo Calculations}
\label{subsec:results_aigk}
\begin{figure}[h!]
    \centering
    \includegraphics{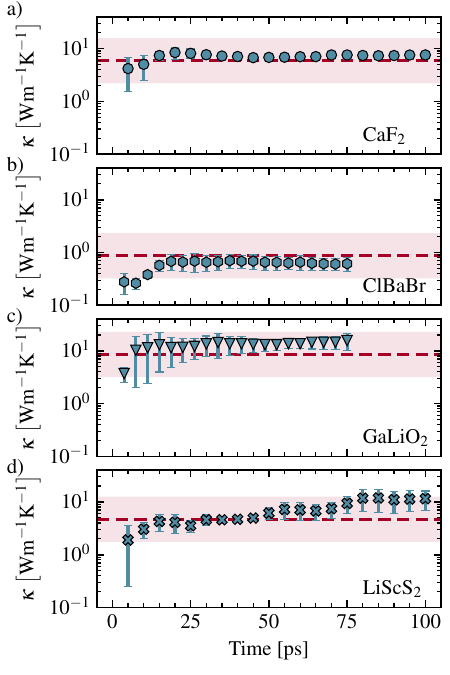}
    \caption{Validation of the predictions of the model. The convergence of the calculated thermal conductivity of a) CaF$_2$, b) ClBaBr, c) GaLiO$_2$ d) LiScS$_2$. All aiGK calculations were done using the average of three 75 ps (ClBaBr and GaLiO$_2$) or 100 ps (CaF$_2$ and LiScS$_2$) molecular dynamics trajectories. The dashed lines are the values of the thermal conductivities predicted by Equation~\ref{eqn:model} and the shaded region is the 95\% confidence interval of the prediction based on the RMSE obtained in Figure~\ref{fig:sisso_training}b.}
    \label{fig:aiGK}
\end{figure}

To confirm that the discovered models produce physically meaningful predictions, we validate the estimated thermal conductivity of four materials using the \textit{ab initio} Green-Kubo method (aiGK)~\cite{Carbogno2017,Knoop2022}.
This approach has recently been demonstrated to be highly accurate when compared to experiments~\cite{Knoop2022}, using similar DFT settings for what was done in this work.
In particular aiGK is highly accurate in the low thermal conductivity regime that we are studying here.
For details of how we calculate $\kappa_\mathrm{L}$ see the methodology in Section~\ref{subsec:methods_aigk}.
For this purpose, we chose BrBaCl, LiScS$_2$, CaF$_2$, and GaLiO$_2$, since these materials represent a broad region of the relevant feature space that also test the boundary regions of the heuristics found by the sensitivity analysis and mapping, as demonstrated by the yellow stars in Figure~\ref{fig:maps}.
Figure~\ref{fig:aiGK} shows the convergence of the thermal conductivity of the selected materials, as calculated from three aiMD trajectories.
All of the calculated thermal conductivities fall within the 95\% confidence interval of the model, with the predictions for both CaF$_2$ and ClBaBr being especially accurate.
The better performance of the model for these materials is expected, as they are more similar to the training data than the hexagonal Caswellsilverite like materials.
Additionally, quantum nuclear effects play a more important role in LiScS$_2$ and GaLiO$_2$ than CaF$_2$ and ClBaBr, which can also explain why those predictions are worse than CaF$_2$ and ClBaBr.
Overall these results demonstrate the predictive power of the discussed model.

\subsection{Discovering Improved Thermal Insulators}
\label{subsec:results_pred}

Using the information gained from the sensitivity analysis and statistical maps of the model, we are now able to design a hierarchical and efficient high-throughput screening protocol split into three stages: structure optimization, harmonic model generation, and anharmonicity quantification.
We demonstrate this procedure by identifying possible thermal insulators within a set of 732 materials, within those compounds available in the materials project~\cite{Jain2013a} that feature the same crystallographic prototypes~\cite{Mehl2017,Hicks2018} as the ones used for training.
Once the geometry is optimized we remove all materials with $V_\mathrm{m} < 35.5$ \AA (60 materials) and all (almost) metallic materials (bandgap $<0.2$~eV), and are left with 302 candidate compounds.
We then generate the converged harmonic model for the remaining materials and screen out all materials with $\Theta_\mathrm{D,\infty} > 547$~K or have an unreliable harmonic model, e.g. materials with imaginary harmonic modes, leaving 148~candidates.
Finally we evaluate the anharmonicity, $\sigma^\mathrm{A}$, for the remaining materials~(see Section~\ref{subsec:methods_sisso_primary_features}) and exclude all materials with $\sigma^\mathrm{A} < 0.206$, and obtain 110 candidate thermal insulators.
To avoid unnecessary calculations, we first estimate $\sigma^\mathrm{A}$ via $\sigma^\mathrm{A}_\mathrm{OS}$ and then refine it via aiMD when $\sigma^\mathrm{A}_\mathrm{OS}>0.4$~\cite{Knoop2020}.
For these candidate materials, we evaluate $\kappa^\mathrm{SISSO}\left(\mathrm{300~K}\right)$ using Eq.~\ref{eqn:model}.
Of the 110 materials that passed all checks, 96 are predicted to have have a $\kappa^\mathrm{SISSO}\left(\mathrm{300~K}\right)$ below 10 Wm$^{-1}$K$^{-1}$, illustrating the success of this method.

\begin{figure}
    \centering
    \includegraphics{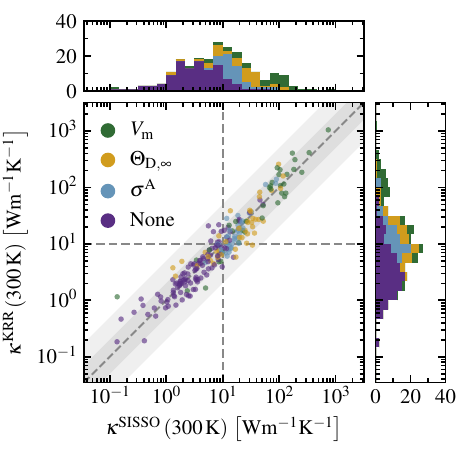}
    \caption{A scatter plot of the prediction of both the SISSO and KRR generated models for an additional 227 materials from the same classes as the training set. $\sigma^\mathrm{A}$ is estimated via $\sigma^\mathrm{A}_\mathrm{OS}$ for all materials with a $\sigma^\mathrm{A}_\mathrm{OS} \leq 0.4$ in this screening. The dataset is split up into four subsets based on if the $V_\mathrm{m}$ test failed (top, green), $\Theta_\mathrm{D,\infty}$ test failed (second from top, yellow), $\sigma^\mathrm{A}$ test failed (third from top, blue), or none of the tests failed (bottom, purple). The outlets correspond to the histogram of all predictions using the same break down. The darker shaded region represents where both predictions are within a factor of 2 of each other and the lighter shaded region where both predictions are within a factor of 5 of each other.}
    \label{fig:prediction}
\end{figure}

Finally, let us emphasize that the proposed strategy is not limited to the discovery of thermal insulators, but can be equally used to find, e.g., good thermal conductors.
This is demonstrated in Figure~\ref{fig:prediction}, in which we predict the thermal conductivity of all non-metallic and stable materials using the SISSO and KRR models.
Generally, both  the SISSO and KRR models agree with each other with only 28 of the 227 materials having a disagreement larger than a factor of two and one (LiHF$_2$) with a disagreement larger than a factor of 5, further illustrating the reliability of these predictions.
We expect that the large deviation for LiHF$_2$ is a result of the large $\sigma^\mathrm{A}$ value for that material (0.54), which is significantly larger than the maximum in the training data.
We can see from the outset histograms of both models that the hierarchical procedure successfully finds the good thermal insulators, with only 26 of the 122 materials with a $\kappa_\mathrm{L}\left(\mathrm{300~K}\right) \leq 10$ Wm$^{-1}$K$^{-1}$ and 10 of the 80 materials with a $\kappa_\mathrm{L}\left(\mathrm{300~K}\right) \leq 5$ Wm$^{-1}$K$^{-1}$ not passing all tests.
Of these eight only the thermal insulating behavior of CuLiF$_2$ and Sr$_2$HN can not be described by the values of the other two tests that passed.
Conversely, materials that do not pass the test show high conductivities.
When one of the tests fail the average estimated value of $\log\left(\kappa_\mathrm{L}\left(\mathrm{300~K}\right)\right)$ increases  to $1.38\pm0.490$ (24.0 Wm$^{-1}$K$^{-1}$), with a range of 0.95 Wm$^{-1}$K$^{-1}$ to 741.3 Wm$^{-1}$K$^{-1}$.
In particular, screening the materials by their molar volumes alone is a good marker for finding strong thermal conductors as all of the 15 materials with $\kappa_\mathrm{L}\left(\mathrm{300~K}\right) \geq 100$ Wm$^{-1}$K$^{-1}$ have a $V_\mathrm{m} \leq 45$ \AA$^3$.

\section{Discussion}
\label{sec:discussion}
We have developed an AI framework to facilitate and accelerate material space exploration, and demonstrate its capabilities for the urgent problem of finding thermal insulators.
By combining symbolic regression and sensitivity analysis, we are able to obtain accurate predictions for a given property using relatively easy to calculate materials properties, while retaining strong physical interpretability.
Most importantly, this analysis enables us to create hierarchical, high-throughput frameworks, which we used to screen over a set of more than 700 materials and find a group of $\sim$100 possible thermal insulators.
Notably, almost all of the good thermal conductors in the set of candidate materials are discarded within the first iteration of the screening, in which we only discriminate by molar volume,~i.e.,~with an absolutely negligible computational cost compared to full calculations of $\kappa_\mathrm{L}$.
Accordingly, we expect this approach to be extremely useful in a wide range of materials problems beyond thermal transport, especially whenever (i)~few reliable data are available, (ii)~additional data are hard to produce, and/or (iii) multiple physical mechanisms compete non-trivially, limiting the reliability of simplified models.

Although the proposed approach is already reliable for small dataset sizes, it obviously becomes more so when applied to larger ones.
Here, the identified heuristics can substantially help steer data creation towards more interesting parts of material space.
Along these lines, it is possible to iteratively refine both the SISSO model and the rules from the sensitivity analysis during material space exploration while the dataset grows.
Furthermore, one can also apply the proposed procedure to the most influential primary features in a recursive fashion, learning new expressions for the computationally expensive features, e.g. $\sigma^\mathrm{A}$, using simpler properties.
In turn, this will further accelerate material discovery, but also allow for gaining further physical insights.
Most importantly, this method is not limited to just the thermal conductivity of a material, and can be applied to any target property.
Further extending this framework to include information about where the underlying electronic structure calculations are expected to fail, also provides a means of accelerating materials discovery more generally~\cite{Duan2021}.

\FloatBarrier

\section{Methods}
\label{sec:method}
\subsection{SISSO}
\label{subsec:methods_sisso}
We use SISSO to discover analytical expressions for $\kappa_\mathrm{L}\left(\mathrm{300~K}\right)$~\cite{Ouyang2017}.
SISSO finds low-dimensional, analytic expressions for a target property, $P$, by first generating an exhaustive set of candidate features, $\hat{\Phi}$, for a given set of primary features, $\hat{\Phi}_0$, and operators $\hat{\mathcal{H}}_m$, and then performing an $\ell_0$-regularization over a subset of those features to find the  $n$-dimensional subset of features, whose linear combination results in the most descriptive model.
$\hat{\Phi}$ is recursively built in rungs, $\hat{\mathcal{F}}_r$, from $\hat{\Phi}_0$ and $\hat{\mathcal{H}}_m$, by applying all elements, $\hat{\mathrm{h}}^m$, of $\mathcal{\hat{H}}^m$ on all elements $\hat{f}_i$ and $\hat{f}_j$ of $\hat{\mathcal{F}}_{r-1}$
\begin{equation*}
    \hat{\mathcal{F}}_r \equiv \hat{\text{h}}^m\left[\hat{f}_i,\hat{f}_j\right], \forall \; \hat{\text{h}}^m \in \mathcal{\hat{H}}^m \;\text{and}\; \forall \; \hat{f}_i,\hat{f}_j \in \hat{\mathcal{F}}_{r-1}.
\end{equation*}
$\hat{\Phi}_r$ is then the union of $\hat{\Phi}_{r-1}$ and $\hat{\mathcal{F}}_r$.
Once $\hat{\Phi}$ is generated, the $n_\mathrm{SIS}$ features most correlated to $P$ are stored in $\mathcal{\hat{S}}_1$, and the best one-dimensional models are trivially extracted from the top elements of $\hat{\mathcal{S}}_1$.
Then the $n_\mathrm{SIS}$ features most correlated to any of the residuals, $\bm{\Delta}_1^i$, of the $n_\mathrm{res}$ best one-dimensional descriptors are stored in $\mathcal{\hat{S}}_2$.
We define this projection  as
\begin{align}
    s   &= \mathrm{max}\left(s_0, s_1, ..., s_i, ...,s_\mathrm{n_\mathrm{res}}\right) \\
    s_i &= R^2\left(\hat{\phi}, \bm{\Delta}_1^i\right),
\end{align}
where $\hat{\phi} \in \hat{\Phi}$, and $R^2$ is the Pearson correlation function.
We call this approach the multiple residual approach, which was first introduced by the authors~\cite{Foppa2021} and later fully described in Ref.~\cite{purcell2023sisso}.
From here, the best two dimensional models are found by performing an $\ell_0$-regularized optimization over $\mathcal{\hat{S}}_1 \cup \mathcal{\hat{S}}_2$ ~\cite{Ghiringhelli2017}.
This process is iteratively repeated until the best $n$-dimensional descriptor is found~\cite{Ouyang2017}.

For this application $\hat{\mathcal{H}}_m$ contains: $A+B$, $A-B$, $A*B$, $\frac{A}{B}$, $\left|A-B\right|$, $\left|A\right|$, $\left(A\right)^{-1}$, $\left(A\right)^{2}$, $\left(A\right)^{3}$, $\sqrt{A}$, $\sqrt[3]{A}$, $\exp\left(A\right)$, $\exp\left(-1.0 * A\right)$, and $\ln\left(A\right)$.
Additionally to ensure the units of the primary features do not affect the final results, we additionally include the following operators: $\left(A + \beta\right)^{-1}$, $\left(A + \beta\right)^{2}$, $\left(A + \beta\right)^{3}$, $\sqrt{\alpha A + \beta}$, $\sqrt[3]{A + \beta}$, $\exp\left(\alpha A\right)$, $\exp\left(-1.0 * \alpha A\right)$, and $\ln\left(\alpha A + \beta\right)$, where $\alpha$ and $\beta$ are scaling and bias constants used to adjust the input data on the fly.
We find the optimal $\alpha$ and $\beta$ terms using non-linear optimization for each of these operators~\cite{Purcell2022,NLopt,purcell2023sisso}.
To ensure that the parameterization does not result in mathematically invalid equations for data points outside of the training set, the range of each candidate feature is derived from the range of the primary features, and the upper and lower bounds for the features are set appropriately.
When generating new expressions these ranges are then used as a domain for the operations, and any expression that would lead to invalid results are excluded~\cite{purcell2023sisso}.
The range of the primary features are set to be physically relevant for the systems we are studying and are listed in Table~\ref{tab:Primary Features}.
Hereafter, we call the use of these operators parametric SISSO.
For more information please refer to~\cite{purcell2023sisso}.

All hyperparameters were set following the cross-validation procedures described in Section~\ref{subsec:methods_error}.

\subsection{Kernel-Ridge Regression}
\label{subsec:methods_krr}

To generate the kernel-ridge regression models we used the utilities provided by scikit-learn~\cite{scikit-learn}, using a radial basis function kernel with optimized regularization term and kernel length scale.
The hyperparameters were selected using with a 141 by 141 point logarithmic grid search with possible parameters ranging from $10^{-7}$ to $10^{0}$.
Before performing the analysis each input feature, $\mathbf{x}_\mathrm{i}$ is standardized
\begin{equation}
    \mathbf{x_\mathrm{i}^\mathrm{Stand}} = \frac{\mathbf{x}_\mathrm{i} - \mu_\mathrm{i}}{\sigma_\mathrm{i}}
\end{equation}
where $\mathbf{x_\mathrm{i}^\mathrm{stand}}$ is the standardized input feature, $\mu_\mathrm{i}$ is the mean of the input feature for the training data, and $\sigma_\mathrm{i}$ is the standard deviation of the input feature for the training data.

\subsection{Gaussian Process Regression}
\label{subsec:methods_gpr}

To generate the Gaussian Process Regression Models we used the utilities provided by scikit-learn~\cite{scikit-learn}, using a radial basis function kernel with an optimized regularization term and kernel length scale.
The hyperparameters were selected using with a 141 by 141 point logarithmic grid search with possible parameters ranging from $10^{-7}$ to $10^{0}$.
Before performing the analysis each input feature, $\mathbf{x}_\mathrm{i}$ is standardized
\begin{equation}
    \mathbf{x_\mathrm{i}^\mathrm{Stand}} = \frac{\mathbf{x}_\mathrm{i} - \mu_\mathrm{i}}{\sigma_\mathrm{i}}
\end{equation}
where $\mathbf{x_\mathrm{i}^\mathrm{stand}}$ is the standardized input feature, $\mu_\mathrm{i}$ is the mean of the input feature for the training data, and $\sigma_\mathrm{i}$ is the standard deviation of the input feature for the training data.
All uncertainty values were taken from the results of the GPR predictions, and in the case of the nested cross-validation the uncertainty was propagated using
\begin{align}
    \kappa^\mathrm{pred}_\mathrm{GPR} &= \frac{1}{3} \sum_{i=1}^3 \kappa^\mathrm{pred}_{\mathrm{GPR},i} \\
    \sigma_{\mathrm{GPR}}^\mathrm{pred} &= \frac{1}{3} \sqrt{\sum_{i=1}^3 \left(\sigma_{\mathrm{GPR}, i}^\mathrm{pred}\right)^2},
\end{align}
where $\kappa^\mathrm{pred}_{\mathrm{GPR}, i}$ and $\kappa^\mathrm{pred}_{\mathrm{GPR}, i}$ are the respective prediction and uncertainty of the $i^{th}$ GPR model for a given data point and $\kappa^\mathrm{pred}_{\mathrm{GPR}}$ and $\kappa^\mathrm{pred}_{\mathrm{GPR}}$ are the respective mean prediction and uncertainty for a prediction.

\subsection{Creating the Dataset}
\label{subsec:methods_sisso_primary_features}
In this study we focus on only room-temperature data for $\kappa_\mathrm{L}$, since values for other temperatures are even scarcer.
However, we note that an explicit temperature dependence can be straightforwardly included using multi-task SISSO~\cite{Ouyang2019}, and it is at least partially included via, the anharmonicity factor,  $\sigma^\mathrm{A}$~\cite{Knoop2020} (see below for more details).
For~$\kappa_\mathrm{L}\left(\mathrm{300~K}\right)$, we have compiled a list of seventy-five materials from the literature (see Supplementary Table~\ref{tab:si_kappa_exp} for complete list with references), whose thermal conductivity has been experimentally measured.
This list was curated from an initial set of over 100 materials, from which we removed all samples that are either thermodynamically unstable or are electrical conductors.
This list of materials covers a diverse set of fourteen different binary and ternary crystal structure prototypes~\cite{Hicks2021,Mehl2017,Hicks2018}.

With respect to the primary features, $\hat{\Phi}_0$, compound specific properties are provided for each material.
All primary features  can be roughly categorized in two classes: Structural parameters that describe the equilibrium structure and dynamical parameters that characterize the nuclear motion.
For the latter case, both harmonic and anharmonic properties have been taken into account.
As shown in~\ref{sec:si_slack_training}, additional features, such as the parameters entering the Slack model, i.e., $\gamma$, $\Theta_\mathrm{a}$, and $V_\mathrm{a}$, can be included.
However, these features do not benefit the model and when included only $V_\mathrm{a}$, and not $\gamma$ or $\Theta_\mathrm{a}$ are selected.
For a complete list of all primary features, and their definitions refer to Table~\ref{tab:Primary Features}.

\begin{table*}
\centering
\caption{List of the primary features used in this calculation}
\label{tab:Primary Features}
\begin{tabular}{D c c c}
\toprule
Name                            & Symbol  & Unit & Domain\\
\hline
Anharmonicity Score (aiMD)~\cite{Knoop2020}      & $\sigma^\mathrm{A}$           & --- & [0.075, 1.0]\\
\hline
Anharmonicity Score (one-shot~\cite{Knoop2020})  & $\sigma^\mathrm{A}_\mathrm{OS}$         & ---  & [0.075, 1.0]\\
\hline
\begin{tabular}{@{}l@{}}Maximum Phonon Frequency \\at the $\Gamma$-point \end{tabular} & $\omega_\mathrm{\Gamma, max}$ &  THz & [0.1, 200] \\
\hline
\begin{tabular}{@{}l@{}}High-Temperature Limit of \\the Debye Temperature \end{tabular} & $\Theta_\mathrm{D,\infty}$ & K & [10,  $1\,000$] \\
\hline
Average Phonon Temperature      & $\Theta_\mathrm{P}$ & K & [10,  $10\,000$] \\
\hline
Heat Capacity                   & $C_\mathrm{V}$      & J mol$^{-1}$ K$^{-1}$ & [10, $5\,000$] \\
\hline
Speed of sound                  & $v_\mathrm{s}$      & m s$^{-1}$ & [500, $10\,000$] \\
\hline
Density                         & $\rho$              & Da \AA$^{-3}$ & [0.25, 10] \\
\hline
Molar Volume                    & $V_\mathrm{m}$      & \AA$^{3}$ & [2.5, $1\,000$] \\
\hline
Minimum Lattice Parameter       & $L_\mathrm{min}$    & \AA & [1, 100] \\
\hline
Maximum Lattice Parameter       & $L_\mathrm{max}$    & \AA & [1, 100] \\
\hline
Mean Lattice Parameter          & $L_\mathrm{avg}$    & \AA & [1, 100] \\
\hline
Reduced Mass                    & $\mu$               & Da & [0.2, 300] \\
\hline
Minimum Atomic Mass             & $m_\mathrm{min}$    & Da & [1, 300] \\
\hline
Maximum Atomic Mass             & $m_\mathrm{max}$    & Da & [1, 300] \\
\hline
Mean Atomic Mass                & $m_\mathrm{avg}$    & Da & [1, 300] \\
\hline
Number of Atoms                 & $n_\mathrm{at}$     & $\mathbb{Z}$ & [1, $1\,000$] \\
\botrule
\end{tabular}
\end{table*}

The structural parameters relate to either the mass of the atoms ($\mu$, $m_\mathrm{min}$, $m_\mathrm{max}$, $m_\mathrm{avg}$), the lattice parameters of the primitive cell ($V_\mathrm{m}$, $L_\mathrm{min}$, $L_\mathrm{max}$, $L_\mathrm{avg}$), the density of the materials ($\rho$), or the number of atoms in the primitive cell ($n_\mathrm{at}$).
For all systems a generalization of the reduced mass, $\mu$, is used so it can be extended to non-binary systems,
\begin{equation}
    \mu = \left(\sum_i^{n_\mathrm{emp}} \frac{1}{m_i} \right)^{-1},
\end{equation}
where $n_\mathrm{emp}$ is the number of atoms in the empirical formula and $m_i$ is the mass of atom, $i$.
Similarly, the molar volume, $V_\mathrm{m}$, is calculated by
\begin{equation}
    V_\mathrm{m} = \frac{V_\mathrm{prim}}{Z},
\end{equation}
where $V_\mathrm{prim}$ is the volume of the primitive cell and $Z = \frac{n_\mathrm{at}}{n_\mathrm{emp}}$.
Finally, $\rho$ is calculated by dividing the total mass of the empirical cell by $V_\mathrm{m}$
\begin{equation}
    \rho = \sum_i^{n_\mathrm{emp}}\frac{m_i}{V_\mathrm{m}}.
\end{equation}

All of the harmonic properties used in these models are calculated from a converged harmonic model generated using phonopy~\cite{phonopy}.
For each material, the phonon density of states of successively larger supercells are compared using a Tanimoto similarity measure
\begin{equation}
    S = \frac{g_\mathrm{p,L}\left(\omega\right) \cdot g_\mathrm{p,S}\left(\omega\right)}{\lVert g_\mathrm{p,L}\left(\omega\right)\rVert^2 + \lVert g_\mathrm{p,S}\left(\omega\right)\rVert^2 - g_\mathrm{p,L}\left(\omega\right) \cdot g_\mathrm{p,S}\left(\omega\right)},
\end{equation}
where $S$ is the similarity score, $g_\mathrm{p,L}\left(\omega\right)$ is the phonon density of states of the larger supercell, $g_\mathrm{p,S}\left(\omega\right)$ is the phonon density of states of the smaller supercell, $A\left(\omega\right)\cdot B\left(\omega\right) = \int_0^\infty A\left(\omega\right)B\left(\omega\right)d\omega$, and $\lVert A\left(\omega\right) \rVert^2=\int_0^\infty A^2\left(\omega\right) d\omega$.
If $S>0.80$, then the harmonic model is considered converged.
From here $C_\mathrm{V}$ is calculated from phonopy as a weighted sum over the mode dependent heat capacities.
Both approximations to the Debye temperature are calculated from the moments of the phonon density of states
\begin{align}
    \langle \varepsilon^n\rangle &= \frac{\int\! d\varepsilon \, g_\mathrm{p}(\varepsilon) \, \varepsilon^n}{\int d\varepsilon g_\mathrm{p}\left(\varepsilon\right)}\\
    \Theta_\mathrm{P} &= \frac{1}{k_B}\langle \varepsilon\rangle\\
    \Theta_\mathrm{D,\infty} &= \frac{1}{k_B}\sqrt{\frac{5}{3} \langle \varepsilon^2\rangle},
\end{align}
where $g_p\left(\varepsilon\right)$ is the phonon density of states at energy $\varepsilon$~\cite{Passler2007}.
Finally $v_\mathrm{s}$ is approximated from the Debye frequency, $\omega_\mathrm{D}$, by~\cite{Toberer2011}
\begin{equation}
    v_\mathrm{s} = \left(\frac{V_\mathrm{a}}{6\pi^2}\right)^{1/3}\omega_\mathrm{D},
\end{equation}
where $\omega_\mathrm{D}$ is approximated as
\begin{equation}
    \omega_\mathrm{D} = \sqrt[3]{\frac{9 n_\mathrm{at}}{a}}
\end{equation}
and $a$ is found by fitting $g_p\left(\omega\right)$ in the range $\left[0, \frac{\omega_\mathrm{\Gamma,max}}{8}\right]$ to
\begin{equation}
    g_{p,D}\left(\omega\right) = a \omega^2.
\end{equation}

To measure the anharmonicity of the materials we use $\sigma^\mathrm{A}$ as defined in ~\cite{Knoop2020}
\begin{eqnarray}
\sigma^\mathrm{A} (T)
&=&\sqrt{\frac{ \sum\limits_{I, \alpha} \left\langle \left( F_{I, \alpha} - F_{I, \alpha}^{\mathrm{ha}} \right)^2 \right\rangle_{(T)}}{
        \sum_{I, \alpha} \left\langle F^2_{I, \alpha} \right\rangle_{(T)}}} ~,
\label{eqn:sigmaAb}
\end{eqnarray}in which $\langle\cdot\rangle_{(T)}$ denotes the thermodynamic average at a temperature $T$, $F_{I, \alpha}$ is the $\alpha$ component of the force calculated from density functional theory (DFT) acting on atom $I$, and $F^\mathrm{ha}_{I,\alpha}$ is the same force approximated by the harmonic model~\cite{Knoop2020}.
First we calculate $\sigma^\mathrm{A}_\mathrm{OS}$, which uses an approximation to the thermodynamic ensemble average using the one-shot method proposed by Zacharias and Giustino~\cite{Zacharias2016}.
In the one-shot approach the atomic positions are offset from their equilibrium positions by a vector $\mathbf{\Delta R}$,
\begin{equation}
\Delta R_{I}^\alpha =  \frac{1}{\sqrt{M_{I}}} \sum_{s} \zeta_s \left\langle {A_s} \right\rangle e_{s I}^{\alpha}~,
\label{eqn:samples1}
\end{equation}
where $I$ is the atom number, $\alpha$ is the component,  $\mathbf e_s$ are the harmonic eigenvectors, $\left\langle {A_s} \right\rangle = {\sqrt{2 k_B T}}/{\omega_s}$ is the mean mode amplitude in the classical limit~\cite{Dove}, and  $\zeta_s = (-1)^{s-1}$~\cite{Zacharias2016}.
These displacements correspond to the turning-points of the oscillation estimated from the harmonic force constants, and is a good approximation to $\sigma^\mathrm{A}$ in the harmonic limit.
Because of this, if $\sigma^\mathrm{A}_\mathrm{OS} < 0.2$ we accept that value as the true $\sigma^\mathrm{A}$.
Otherwise we calculate $\sigma^\mathrm{A}$ using aiMD in the canonical ensemble at 300 K for 10 ps, using the Langevin thermostat.
When performing the high-throughput screening the threshold for when to use aiMD is increased to 0.4 because that is the point that $\sigma^\mathrm{A}_\mathrm{OS}$ becomes qualitatively unreliable~\cite{Knoop2020}.

All electronic structure calculations are done using FHI-aims~\cite{FHI-aims}.
All geometries are optimized with symmetry-preserving, parametric constraints until all forces are converged to a numerical precision better than \mbox{10$^{-3}$\,eV/\AA}~\cite{Lenz2019}.
The constraints are generated using the AFlow XtalFinder Tool~\cite{Hicks2021}.
All calculations use the PBEsol functional to calculate the exchange-correlation energy and an SCF convergence criteria of $10^{-6}$ eV/\AA~and $5\times10^{-4}$ eV/\AA~for the density and forces, respectively.
Relativistic effects are included in terms of the scalar atomic ZORA approach and all other settings are taken to be the default values in FHI-aims.
For all calculations we use the \textit{light} basis sets and numerical settings in FHI-aims.
These settings were shown to ensure a convergence in lattice constants of~$\pm 0.1~\mbox{\AA}$ and a relative accuracy in phonon frequencies of~3\%~\cite{Knoop2020}.

All primary features are calculated using the workflows defined in FHI-vibes~\cite{Knoop2020a}.

\subsection{Error Evaluation}
\label{subsec:methods_error}

To estimate the prediction error for all models we perform a nested cross-validation, where the data are initially separated into different training and test sets using a ten-fold split.
Two hyperparameters (maximum dimension and parameterization depth) are then optimized using a five-fold cross validation on each of the training sets, and the overall performance of the model is evaluated on the corresponding test set.
The size of the SIS subspace, number of residuals, and rung were all set to $2\,000$, 10, and 3, respectively, because they did not have a large impact on the final results.
We then repeat the procedure three times and average over each iteration to get a reliable estimate of the prediction error for each sample~\cite{Krstajic2014}.

\subsection{Calculating the inputs to the Slack model}
\label{subsec:methods_slack}

The individual components for the Slack model were the same as the ones used for the main models, with the exception of $\gamma$, $V_\mathrm{a}$ and $\Theta_\mathrm{a}$.
For $\Theta_\mathrm{a}$, we first calculate the Debye temperature, $\Theta_\mathrm{D}$
\begin{equation}
    \Theta_\mathrm{D} = \frac{\hbar \omega_\mathrm{D}}{k_B}
\end{equation}
where $\omega_\mathrm{D}$ is the same Debye frequency used for calculating $v_\mathrm{s}$ (see Section~\ref{subsec:methods_sisso_primary_features}), $k_B$ is the Boltzmann constant, and $\hbar$ is Planck's constant.
From here we calculate $\Theta_\mathrm{a}$ using
\begin{equation}
    \Theta_\mathrm{a} = \frac{\Theta_\mathrm{D}}{\sqrt[3]{n_\mathrm{at}}}.
\end{equation}
We use the phonopy definition of $\Theta_\mathrm{D}$ instead of $\Theta_\mathrm{D,\infty}$ because it is better aligned to the original definition of $\Theta_\mathrm{a}$.
However, it is not used in the SISSO training because the initial fitting procedure to find $\omega_\mathrm{D}$ does not produce a unique value for $\Theta_\mathrm{D}$ and it is already partially included via $v_\mathrm{s}$.
To calculate the thermodynamic Gr\"uneisen parameter we use the utilities provided by phonopy~\cite{phonopy}.
The atomic volume was calculated by taking the volume of the primitive cell and dividing it by the total number of atoms.

\subsection{Calculating the Sobol Indexes}
\label{subsec:methods_sobol}

Formally, the Sobol indices are defined as
\begin{align}
    S_i     &=     \frac{\mathrm{Var}_{\hat{x}_i}               \left(E_{\widetilde{\mathcal{X}}_{i}}\left(\log\left(\kappa_\mathrm{L}\left(\mathrm{300~K}\right)\right) | \hat{x}_i                     \right) \right)}{\mathrm{Var}\left(\log\left(\kappa_\mathrm{L}\left(\mathrm{300~K}\right)\right)\right)} \\
    S_i^\mathrm{T} &= 1 - \frac{\mathrm{Var}_{\widetilde{\mathcal{X}}_i}\left(E_{\hat{x}_i}                      \left(\log\left(\kappa_\mathrm{L}\left(\mathrm{300~K}\right)\right) | \widetilde{\mathcal{X}}_i\right) \right)}{\mathrm{Var}\left(\log\left(\kappa_\mathrm{L}\left(\mathrm{300~K}\right)\right)\right)}
\end{align}
where $\hat{x}_i \in \hat{\mathcal{X}}$ is one of the inputs to the model,  $\mathrm{Var}_{a}\left(B\right)$ is the variance of $B$ with respect to $a$, $E_a\left(B\right)$ is the mean of $B$ after sampling over $a$, and $\widetilde{\mathcal{X}}_i$ is the set of all variables excluding $\hat{x}_i$.

Normally, it is assumed that all elements of $\hat{\mathcal{X}}$ are independent of each other, and this assumption is preserved when calculating $S_i$ and $S_i^T$ in Figure~\ref{fig:sensitivity_analysis}b.
As a result of this, the variance of $\log\left(\kappa^\mathrm{SISSO}\left(\mathrm{300~K}\right)\right)$ and the required expectation values would be calculated from sampling over an $n_v$-dimensional hypercube covering the full input range, ignoring the correlation between the input variables.
However, in order to properly model the correlative effects between elements of $\hat{\mathcal{X}}$, Kucherenko et al. modify this sampling approach~\cite{Kucherenko2012,Wiederkehr2018}.
The first step of the updated algorithm is to fit the input data to a set of marginal univariate distributions coupled together via a copula~\cite{Kucherenko2012,Wiederkehr2018}.
The algorithm then samples over an $n_v$-dimensional unit-hypercube and transforms these samples into the correct variable space using a transform defined by the fitted distributions and copulas (see~\ref{sec:si_sa} for more details).
It was later demonstrated that when using the approach proposed by Kucherenko and coworkers to calculate the Sobol indices, $S_i$ includes effects from the dependence of $\hat{x}_i$ on those in $\widetilde{\mathcal{X}}_{i}$, while $S^\mathrm{T}_i$ is independent of these effects~\cite{Mara2015a}.
We use this updated algorithm to calculate $S_i$ and $S_i^\mathrm{T}$ in Figure~\ref{fig:sensitivity_analysis}a.
In both cases we use the implementation in UQLab~\cite{Marelli2014} to calculate $S_i$ and $S_i^T$.

\subsection{Calculating the SHAP Indexes}
\label{subsec:methods_shap}
The SHAP values are calculated by treating the features as independent variables using the original method proposed by Lundberg and Lee~\cite{NIPS2017_7062}, as implemented in the python package \textsc{shap}, and as dependent variables using \textsc{shapr} by Aas, \textit{et al.}~\cite{Aas2021}.
The SHAP values are an extension of the Shapley values from cooperative game theory, that distributes the contribution, $v\left(\mathcal{S}\right)$, of each player or subset of players, $\mathcal{S} \subseteq \mathcal{M}=\left\{1, \cdot, M\right\}$, where $\mathcal{M}$ is the set of all players~\cite{Aas2021,NIPS2017_7062}.
The Shapley value, $\phi_j\left(v\right)=\phi_j$, can then be calculated by taking a weighted mean over the contribution function differences for all $\mathcal{S}$ not containing the player, $j$,
\begin{equation}
    \begin{split}
        \phi_j = \sum_{\mathcal{S}\subseteq\mathcal{M} \setminus \left\{j\right\}}& \frac{\left|\mathcal{S}\right|! \left(M - \left|\mathcal{S}\right| - 1\right)!}{M!} \\ &\left(v\left(\mathcal{S}\cup\left\{j\right\} \right) - v\left(\mathcal{S}\right)\right),\\& j=1,\cdots,M,
    \end{split}
\end{equation}
where $\left|\mathcal{S}\right|$ is the number of members in $\mathcal{S}$~\cite{Aas2021}.
For a machine learning problem with a training set $\left\{y^i,\bm{x}^i\right\}_{i=1,\cdots,n_\mathrm{train}}$, where $y^i$ is the property value and $\bm{x}^i$ are the target property value and input feature values for the $i^{th}$ data point in the training set with $n_{train}$ data points~\cite{Aas2021,NIPS2017_7062}, we can explain the prediction of the model, $f\left(\bm{x}^*\right)$ for a particular point, $\bm{x}^*$, with
\begin{equation}
f\left(\bm{x}^*\right) = \phi_0 + \sum_{j=1}^M \phi_j^*,
\end{equation}
where $\phi_0$ is the mean prediction and $\phi_j^*$ is the Shapley value for the $j^{th}$ feature for a prediction $\bm{x}=\bm{x}^*$.
Essentially the Shapley value for the model describes the difference between a prediction, $y^* = f\left(\bm{x}^*\right)$, and the mean of all predictions~\cite{Aas2021,NIPS2017_7062}.
The contribution function is then defined as
\begin{equation}
v\left(\mathcal{S}\right) = E\left[\left.f\left(\bm{x}\right)\right| \bm{x}_\mathcal{S} = \bm{x}_\mathcal{S}^*\right],
\end{equation}
which is the expectation value of the model conditional on $\bm{x}_\mathcal{S} = \bm{x}_\mathcal{S}^*$~\cite{Aas2021,NIPS2017_7062}.
The expectation value can be calculated as
\begin{equation}
\begin{split}
&E\left[\left.f\left(\bm{x}\right)\right| \bm{x}_\mathcal{S} = \bm{x}_\mathcal{S}^*\right] = E\left[\left.f\left(\bm{x}_{\widetilde{\mathcal{S}}} ,\bm{x}_\mathcal{S}\right)\right| \bm{x}_\mathcal{S} = \bm{x}_\mathcal{S}^*\right]\\
&=\int f\left(\bm{x}_{\widetilde{\mathcal{S}}} ,\bm{x}_\mathcal{S}\right)p\left(\left.\bm{x}_{\widetilde{\mathcal{S}}}\right|\bm{x}_\mathcal{S} = \bm{x}_\mathcal{S}*\right)d\bm{x}_{\widetilde{\mathcal{S}}},
\end{split}
\end{equation}
where $\bm{x}_{\widetilde{\mathcal{S}}}$ is the subset of all features not included in $\mathcal{S}$ and $p\left(\left.\bm{x}_{\widetilde{\mathcal{S}}}\right|\bm{x}_\mathcal{S} = \bm{x}_\mathcal{S}*\right)$ is the conditional probability distribution of $\bm{x}_{\widetilde{\mathcal{S}}}$ given $\bm{x}_\mathcal{S} = \bm{x}_\mathcal{S}*$~\cite{Aas2021,NIPS2017_7062}.
In the case where the features are treated independently, $p\left(\left.\bm{x}_{\widetilde{\mathcal{S}}}\right|\bm{x}_\mathcal{S} = \bm{x}_\mathcal{S}*\right)$ is replaced by $p\left(\bm{x}_{\widetilde{\mathcal{S}}}\right)$ and $v\left(\mathcal{S}\right)$ can be approximated by Monte Carlo integration
\begin{equation}
v\left(\mathcal{S}\right) = \frac{1}{K}\sum_{k=1}^Kf\left(\bm{x}_{\widetilde{\mathcal{S}}}^k, \bm{x}_\mathcal{S}^*\right),
\end{equation}
where $\bm{x}_{\widetilde{\mathcal{S}}}^k$ are samples from the training data, and $K$ is the number of samples taken~\cite{Aas2021,NIPS2017_7062}.
To include feature dependence the marginal distributions of the training data are converted into a Gaussian copula and that is used to generate samples for the Monte Carlo integration~\cite{Aas2021}.

Because the number of subsets that need to be explored grows as $2^M$ for the number of features, calculating the exact Shapley values for a large number of inputs becomes intractable.
To remove this constraint the problem can be approximated as the optimal solution of a weighted least squares problem, which can be described as Kernel SHAP, which is described in~\cite{Aas2021,NIPS2017_7062}.

\subsection{Calculating the LIME Indexes}
\label{subsec:methods_lime}
For the LIME values we use the \textsc{lime} package in python~\cite{lime}.
The values were calculated using the standard tabular explainer using all features in the model and the mean absolute value of each prediction for each feature was used to asses the global feature importance.
The methodology assumes the features are independent and for algorithmic details see Ref.~\cite{lime}

\subsection{Calculating the Thermal Conductivity}
\label{subsec:methods_aigk}

To calculate $\kappa_\mathrm{L}$, we use the \textit{ab initio} Green Kubo (aiGK) method~\cite{Carbogno2017, Ravichandran2018}.
The aiGK method calculates the $\alpha\beta$ component of the thermal conductivity tensor, $\kappa^{\alpha\beta}$, of a material for a given volume $V$, pressure $p$, and temperature $T$ with
\begin{equation}
    \kappa^{\alpha\beta}\left(T, p\right) = \frac{V}{k_B T^2}\lim_{\tau \to \infty}\int_0^\tau \langle G\left[\mathbf{J}\right]^{\alpha\beta}\left(\tau^\prime\right)\rangle_{\left(T,p\right)}d\tau^\prime
\end{equation}
where $k_B$ is Boltzmann's constant, $\langle\cdot\rangle_{\left(T,p\right)}$ denotes an ensemble average, $\mathbf{J}\left(t\right)$ is the heat flux, and $G\left[\mathbf{J}\right]$ is the time-(auto)correlation functions
\begin{equation}
    G\left[\mathbf{J}\right]^{\alpha\beta} = \lim_{t_0\to\infty}\frac{1}{t_0}\int_0^{t_0-\tau}J^\alpha\left(t\right)J^\beta\left(t + \tau\right) dt.
\end{equation}
The heat flux of each material is calculated from aiMD trajectories using the following definition
\begin{equation}
    \mathbf{J}\left(t\right)=\sum_I\bm{\sigma}_I \dot{\mathbf{R}}_I,
\end{equation}
where $\mathbf{R}_I$ is the position of the $i^{th}$-atom and $\bm{\sigma}_I$ is the contribution of the $i^{th}$ atom to the stress tensor, $\bm{\sigma}=\sum_I \bm{\sigma}_I$~\cite{Carbogno2017}.
From here $\kappa_\mathrm{L}$ is calculated as
\begin{equation}
    \kappa_\mathrm{L} = \frac{1}{3} \mathrm{Tr}\left[\bm{\kappa}\right]
\end{equation}
All calculations were done using both FHI-vibes~\cite{Knoop2020a} and FHI-aims with the same settings as the previous calculations~\cite{Knoop2020} (see Section~\ref{subsec:methods_sisso_primary_features} for more details).
The molecular dynamics calculations were done using a 5 fs time step in the NVE ensemble, with the initial structures taken from a 10 ps NVT trajectory.
Three MD calculations were done for each material and the $\kappa_\mathrm{L}$ was taken to be the average of all three runs.

\section{Data Availability}
All raw electronic structure data can be found on the NOMAD archive~(\url{https://dx.doi.org/10.17172/NOMAD/2022.04.27-1})~\cite{Purcell2022Data}.
All processed data and figure creation scripts can be found on figshare~(\url{https://doi.org/10.6084/m9.figshare.22068749.v4})~\cite{Purcell2023Figshare}.
A reproduction notebook can be found on the NOMAD AI Toolkit~(\url{https://nomad-lab.eu/aitutorials/kappa-sisso}).

\section{Code Availability}
\label{sec:code_avail}
\textsc{SISSO++}~\cite{Purcell2022} and \textsc{FHI-vibes}~\cite{Knoop2020a} were used to generate all data and analysis in the paper and are freely available online in the cited publications.
All electronic structure calculations were done using \textsc{FHI-aims}~\cite{FHI-aims}, which is freely available for use for academic use (with a voluntary donation) (\url{https://fhi-aims.org/get-the-code-menu/get-the-code}).
The Sobol indexes are calculated with \textsc{UQLab}~\cite{Marelli2014,Wiederkehr2018} (\url{https://www.uqlab.com/download}) and the KERNEL shap values were found with \textsc{shapr}~\cite{Aas2021} (\url{https://github.com/NorskRegnesentral/shapr}) which are open source.
The python SHAP library~\cite{NIPS2017_7062} was also used for the independent SHAP values, and is open source (\url{https://github.com/slundberg/shap}).

\begin{acknowledgments}
T.A.R.P. thanks Florian Knoop for valuable discussions and providing scripts for the \textit{ab initio} Green Kubo analysis.
This work was funded by the NOMAD Center of Excellence (European Union's Horizon 2020 research and innovation program, grant agreement Nº 951786), the ERC Advanced Grant TEC1p (European Research Council, grant agreement Nº 740233), and the project FAIRmat (FAIR Data Infrastructure for Condensed-Matter Physics and the Chemical Physics of Solids, German Research Foundation, project Nº 460197019). T.A.R.P. would like to thank the Alexander von Humboldt (AvH) Foundation for their support through the AvH Postdoctoral Fellowship Program.
This research used resources of the Max Planck Computing and Data Facility and the Argonne Leadership Computing Facility, which is a DOE Office of Science User Facility supported under Contract DE-AC02-06CH11357.
\end{acknowledgments}

\section{Author Contributions}
TARP implemented all methods and performed all calculations. TARP and CC ideated the workflow. MS, LMG and CC supervised the project. All authors analyzed the data and wrote the manuscript.

\section{Competing Interests}
The Authors declare no Competing Financial or Non-Financial Interests.

\clearpage
\onecolumngrid

\section*{Supplementary Information}

\setcounter{table}{0}
\setcounter{figure}{0}
\setcounter{equation}{0}
\renewcommand{\thetable}{\arabic{table}}
\renewcommand{\theequation}{Supplementary \arabic{equation}}
\renewcommand{\thesubsection}{Supplementary Note \arabic{subsection}}
\renewcommand{\tablename}{Supplementary Table}
\renewcommand{\figurename}{Supplementary Figure}

\subsection{Experimental Values of Thermal Conductivity}
\label{sec:si_experimental_kappa}
Supplementary Table~\ref{tab:si_kappa_exp} lists all values of thermal conductivity used for the learning, as well as the references for those values.

\begin{longtable*}{MNMM|MNMM}
\caption{A list of all thermal conductivity values used for training, including references}\\
\hline
&&&&&&&
\label{tab:si_kappa_exp}\\
Material    & AFLOW Prototype            & \begin{tabular}[c]{@{}c@{}}$\kappa_\mathrm{L}$\\(Wm$^{-1}$K$^{-1}$)\end{tabular} & Ref. & Material    & AFLOW Prototype            & \begin{tabular}[c]{@{}c@{}}$\kappa_\mathrm{L}$\\(Wm$^{-1}$K$^{-1}$)\end{tabular} & Ref. \\
\hline
C           & A\_cF8\_227\_a             & 3000                       & \cite{Morelli} &
Ge          & A\_cF8\_227\_a             & 65                         & \cite{Morelli} \\
Si          & A\_cF8\_227\_a             & 166                        & \cite{Morelli} &
BaO         & AB\_cF8\_225\_a\_b         & 2.3                        & \cite{Morelli} \\
CaO         & AB\_cF8\_225\_a\_b         & 30                         & \cite{Morelli} &
KBr         & AB\_cF8\_225\_a\_b         & 3.4                        & \cite{Morelli} \\
KCl         & AB\_cF8\_225\_a\_b         & 7.1                        & \cite{Morelli} &
KF          & AB\_cF8\_225\_a\_b         & 6.43                       & \cite{Morelli} \\
KI          & AB\_cF8\_225\_a\_b         & 2.6                        & \cite{Morelli} &
LiBr        & AB\_cF8\_225\_a\_b         & 1.83                       & \cite{Morelli} \\
LiF         & AB\_cF8\_225\_a\_b         & 17.6                       & \cite{Morelli} &
LiH         & AB\_cF8\_225\_a\_b         & 15                         & \cite{Morelli} \\
MgO         & AB\_cF8\_225\_a\_b         & 60                         & \cite{Morelli} &
NaBr        & AB\_cF8\_225\_a\_b         & 2.8                        & \cite{Morelli} \\
NaCl        & AB\_cF8\_225\_a\_b         & 7.1                        & \cite{Morelli} &
NaF         & AB\_cF8\_225\_a\_b         & 18.4                       & \cite{Morelli} \\
NaI         & AB\_cF8\_225\_a\_b         & 1.8                        & \cite{Morelli} &
PbS         & AB\_cF8\_225\_a\_b         & 2.9                        & \cite{Morelli} \\
PbSe        & AB\_cF8\_225\_a\_b         & 2                          & \cite{Morelli} &
PbTe        & AB\_cF8\_225\_a\_b         & 2.5                        & \cite{Morelli} \\
RbBr        & AB\_cF8\_225\_a\_b         & 3.8                        & \cite{Morelli} &
RbCl        & AB\_cF8\_225\_a\_b         & 2.8                        & \cite{Morelli} \\
RbF         & AB\_cF8\_225\_a\_b         & 2.27                       & \cite{Morelli} &
RbI         & AB\_cF8\_225\_a\_b         & 2.3                        & \cite{Morelli} \\
SrO         & AB\_cF8\_225\_a\_b         & 10                         & \cite{Morelli} &
CsBr        & AB\_cP2\_221\_b\_a         & 0.94                       & \cite{Gerlich1982} \\
CsCl        & AB\_cP2\_221\_b\_a         & 1                          & \cite{Gerlich1982} &
CsI         & AB\_cP2\_221\_b\_a         & 1.1                        & \cite{Gerlich1982} \\
AlAs        & AB\_cF8\_216\_c\_a         & 98                         & \cite{Morelli} &
AlP         & AB\_cF8\_216\_a\_c         & 90                         & \cite{Morelli} \\
AlSb        & AB\_cF8\_216\_a\_c         & 56                         & \cite{Morelli} &
BN          & AB\_cF8\_216\_a\_c         & 760                        & \cite{Morelli} \\
BP          & AB\_cF8\_216\_a\_c         & 350                        & \cite{Morelli} &
CdSe        & AB\_cF8\_216\_a\_c         & 4.4                        & \cite{Morelli} \\
CdTe        & AB\_cF8\_216\_a\_c         & 7.5                        & \cite{Morelli} &
CSi         & AB\_cF8\_216\_c\_a         & 360                        & \cite{Morelli} \\
GaAs        & AB\_cF8\_216\_c\_a         & 45                         & \cite{Morelli} &
GaP         & AB\_cF8\_216\_a\_c         & 100                        & \cite{Morelli} \\
GaSb        & AB\_cF8\_216\_a\_c         & 40                         & \cite{Morelli} &
InAs        & AB\_cF8\_216\_c\_a         & 30                         & \cite{Morelli} \\
InP         & AB\_cF8\_216\_a\_c         & 93                         & \cite{Morelli} &
InSb        & AB\_cF8\_216\_a\_c         & 20                         & \cite{Morelli} \\
ZnS         & AB\_cF8\_216\_c\_a         & 27                         & \cite{Morelli} &
ZnSe        & AB\_cF8\_216\_c\_a         & 19                         & \cite{Morelli} \\
ZnTe        & AB\_cF8\_216\_c\_a         & 18                         & \cite{Morelli} &
AlN         & AB\_hP4\_186\_b\_b         & 350                        & \cite{Morelli} \\
BeO         & AB\_hP4\_186\_b\_b         & 370                        & \cite{Morelli} &
CdS         & AB\_hP4\_186\_b\_b         & 16                         & \cite{Morelli} \\
CSi         & AB\_hP4\_186\_b\_b         & 490                        & \cite{Morelli} &
GaN         & AB\_hP4\_186\_b\_b         & 210                        & \cite{Morelli} \\
ZnO         & AB\_hP4\_186\_b\_b         & 60                         & \cite{Morelli} &
SbCoTi      & ABC\_cF12\_216\_c\_b\_a    & 12                         & \cite{Kawaharada2004} \\
SnNiTi      & ABC\_cF12\_216\_c\_b\_a    & 9.3                        & \cite{Hohl1999} &
VFeSb       & ABC\_cF12\_216\_c\_a\_b    & 13                         & \cite{Young1999} \\
ZrNiSn      & ABC\_cF12\_216\_c\_b\_a    & 8.8                        & \cite{Hohl1999} &
Li$_2$O     & A2B\_cF12\_225\_c\_a       & 11                         & \cite{Takahashi1980} \\
Mg$_2$Ge    & AB2\_cF12\_225\_a\_c       & 9.3                        & \cite{Martin1972} &
Mg$_2$Si    & A2B\_cF12\_225\_c\_a       & 8.2                        & \cite{Martin1972} \\
Mg$_2$Sn    & A2B\_cF12\_225\_c\_a       & 7.1                        & \cite{Martin1972} &
Cu$_2$O     & A2B\_cP6\_224\_b\_a        & 5                          & \cite{Chen2019b} \\
CoSb$_3$    & AB3\_cI32\_204\_c\_g       & 10                         & \cite{Morelli1995} &
Al$_2$O$_3$ & A2B3\_hR10\_167\_c\_e      & 30                         & \cite{Slack1962} \\
Cr$_2$O$_3$ & A2B3\_hR10\_167\_c\_e      & 13                         & \cite{WILLIAMS1984} &
AgGaS$_2$   & ABC2\_tI16\_122\_b\_a\_d   & 1.45                       & \cite{Toher2014b} \\
CdGeP$_2$   & ABC2\_tI16\_122\_a\_b\_d   & 11                         & \cite{Valeri-Gil1993} &
CuGaS$_2$   & ABC2\_tI16\_122\_b\_a\_d   & 5.09                       & \cite{Toher2014b} \\
CuGaTe$_2$  & ABC2\_tI16\_122\_b\_a\_d   & 2.2                        & \cite{Toher2014b} &
CdAs$_2$Ge  & A2BC\_tI16\_122\_d\_b\_a   & 8.32                       & \cite{Huang2016} \\
ZnAs$_2$Ge  & A2B\_cF12\_225\_c\_a       & 11                         & \cite{Valeri-Gil1993} &
ZnAs$_2$Si  & A2BC\_tI16\_122\_d\_b\_a   & 14                         & \cite{Valeri-Gil1993} \\
ZnGeP$_2$   & AB2C\_tI16\_122\_b\_d\_a   & 18                         & \cite{Valeri-Gil1993} &
AlCuO$_2$   & ABC2\_hR4\_166\_b\_a\_c    & 28.05                      & \cite{Lu2015, Pantian2017} \\
Ga$_2$O$_3$ & A2B3\_mC20\_12\_2i\_3i     & 14                         & \cite{Villora2008} &
Sc$_2$O$_3$ & A3B2\_cI80\_206\_e\_bd     & 17                         & \cite{Li2003} \\
SnO$_2$     & A2B\_tP6\_136\_f\_a        & 98                         & \cite{Turkes1980a} & & & & \\
\botrule
\end{longtable*}
\FloatBarrier

\subsection{Predicted Thermal Conductivity from Each Model}
\label{sec:si_parity_comp}
\begin{figure}
    \centering
    \includegraphics{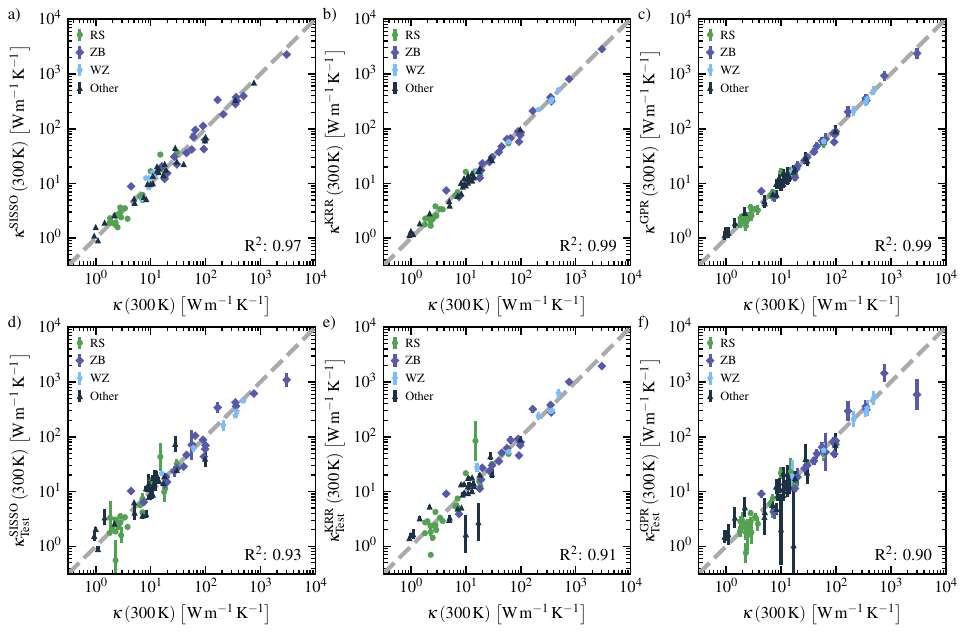}
    \caption{Comparison of the predicted $\kappaRT^\mathrm{pred}$ against the measured $\kappaRT$ for the model trained against all data (a-c) and the average of the three nested cross-validation runs (d-f) for the SISSO (a and d), KRR (b and e) and GPR (c and f) models. The error bars are the standard deviation of either the GPR model (c), the three predictions from the nested cross-validation (d-e), or the propagated uncertainties of the three GPR predictions (f).}
    \label{fig:si_all_parity}
\end{figure}

Supplementary Figure~\ref{fig:si_all_parity} compares the experimental thermal conductivity of each material to the corresponding predicted values of $\kappaRT$ from each model generated from the training and test sets.
When the entire dataset is used in training both KRR and GPR outperform SISSO; however, when averaging over the three predictions for each material from the nested cross-validation study the SISSO model slightly outperforms both KRR and GPR.
The increased error of the KRR and GPR predictions is likely a result of a small subset of materials at the boundaries of the training set, where larger extrapolative errors can occur.
This is best illustrated for Sc$_2$O$_3$ and CoSb$_3$ in Supplementary Figure~\ref{fig:si_all_parity}e and f as the two main outliers.
The SISSO model performs better in these regions as the larger overall uncertainty, as measured by the standard deviation of the three predictions, leads to a possible cancellation of errors.
Outside of this region the uncertainty estimate of the trained GPR model largely matches what is seen during cross-validation, suggesting that the model is reliable when in the interpolative regime.
\FloatBarrier

\subsection{Synthetic Data Generation for the Sensitivity Analysis}
\label{sec:si_sa}

The synthetic data used to perform the Sobol analysis is generated from a multivariate distribution fitted to the training data represented by a series of univariate marginal distributions and a Gaussian copula as summarized in Supplementary Tables~\ref{tab:si_marginal_distribution} and \ref{tab:si_copula}, respectively.
The distributions used are the gamma, log-normal, Rayleigh, Weibull, and uniform distributions.
The probability density function for the gamma distribution is defined as
\begin{equation}
    f(x)= \frac{1}{\Gamma\left(k\right)\theta^k} x^{k-1}e^{-\frac{x}{\theta}}
\end{equation}
The probability density function for the log-normal distribution is defined as
\begin{equation}
    f(x)= \frac{1}{x \sigma \sqrt{2 \pi}} \exp \left(-\frac{(\ln (x)-\mu)^{2}}{2 \sigma^{2}}\right)
\end{equation}
The probability density function for the Rayleigh distribution is defined as
\begin{equation}
    f(x) = \frac{x}{\sigma^{2}} e^{-x^{2} /\left(2 \sigma^{2}\right)}
\end{equation}
The probability density function for the Weibull distribution is defined as
\begin{equation}
    f(x)=\left\{\begin{array}{ll}
\frac{k}{\lambda}\left(\frac{x}{\lambda}\right)^{k-1} e^{-(x / \lambda)^{k}}, & x \geq 0 \\
0, & x<0
\end{array}\right.
\end{equation}
The probability density function for the uniform distribution is defined as
\begin{equation}
    f(x) = \left\{\begin{array}{ll}
\frac{1}{b-a} & \text { for } x \in[a, b] \\
0 & \text { otherwise }
\end{array}\right.
\end{equation}
For all distributions the values of the constants are listed in~\ref{tab:si_marginal_distribution}

\begin{table*}
\centering
\caption{Summary of the univariate marginal distributions used to generate the synthetic data}
\label{tab:si_marginal_distribution}
\begin{tabular}{ADD}
\toprule
  & Type & Parameters\\
\midrule
$\rho$                       & Gamma      & $\theta=0.4077$, $k=6.7042$      \\
$\ThetaD$                    & Log-Normal & $\mu = 6.0163$, $\sigma=0.6575$ \\
$\Vm$                        & Weibull    & $\lambda=58.4857$, $k=2.2902 $ \\
$m_\mathrm{avg}$             & Rayleigh   & $\sigma=48.7896$                  \\
$\sigmaA$                    & Uniform    & $a=0.075$, $b=0.425$              \\
$\omega_\mathrm{\Gamma,max}$ & Log-Normal & $\mu = 2.0993$, $\sigma=0.608$  \\
$\mu$                        & Gamma      & $\theta=12.6303$, $k=1.7522$     \\
\botrule
\end{tabular}
\end{table*}

The Gaussian copula is defined by
\begin{equation}
    C_R^\mathrm{Gauss}\left(u\right) = \Phi_R\left(\Phi^{-1}\left(u_1\right), \ldots, \Phi^{-1}\left(u_d\right) \right),
\end{equation}
where $\Phi^{-1}$ is the inverse cumulative distribution function of a standard normal and $\Phi_R$ is the joint cumulative distribution function of a multivariate normal distribution with a mean zero vector and a covariance matrix equal to the correlation matrix $R$ defined in Supplementary Table~\ref{tab:si_copula}.
\begin{table*}
\centering
\caption{The Pearson correlation parameters used for the Gaussian Copula}
\label{tab:si_copula}
\begin{tabular}{AAAAAAAA}
\toprule
  & $\rho$ & ${\ThetaD}$ & ${\Vm}$ & $m_\mathrm{avg}$ & ${\sigmaA}$ & $\omega_\mathrm{\Gamma,max}$ & $\mu$\\
\midrule
$\rho$                       &  1.0 & -0.2304 &  0.3010 &  0.7008 & -0.0982 & -0.2084 &  0.4605 \\
$\ThetaD$                    & -0.2304 &  1.0 & -0.7316 & -0.8022 & -0.7371 &  0.9549 & -0.7593 \\
$\Vm$                        &  0.3010 & -0.7316 &  1.0 &  0.6976 &  0.4423 & -0.5880 &  0.4442 \\
$m_\mathrm{avg}$             &  0.7008 & -0.8022 &  0.6976 &  1.0 &  0.3850 & -0.7519 &  0.8609 \\
$\sigmaA$                    & -0.0982 & -0.7371 &  0.4423 &  0.3850 &  1.0 & -0.7490 &  0.3019 \\
$\omega_\mathrm{\Gamma,max}$ & -0.2084 &  0.9549 & -0.5880 & -0.7519 & -0.7490 &  1.0 & -0.7133 \\
$\mu$                        &  0.4605 & -0.7593 &  0.4442 &  0.8609 &  0.3019 & -0.7133 &  1.0 \\
\botrule
\end{tabular}
\end{table*}

Using this dataset we are able to map out the models against each of the primary features, and pairs of primary features in Supplementary Figure~\ref{fig:si_all_maps}.
As expected by the correlations shown in Supplementary Table~\ref{tab:si_copula} the maps of $\ThetaD$ and $\omega_\mathrm{\Gamma,\, max}$ are very similar and give the same insights.
For $\rho$, $m_\mathrm{avg}$, and $\mu$ we find that the entire range of possible values would be considered to contain good thermal insulators.
This is likely a result of the weak dependence between each of these variables and $\kappaRTp$ combined with the average value for $\log\left(\kappaRT\right)$ of 1.2 for this dataset leading to relatively flat curves around 10 Wm$^{-1}$K$^{-1}$ for the expected value of $\kappaRTp$.

\begin{figure}[htp!]
    \centering
    \includegraphics{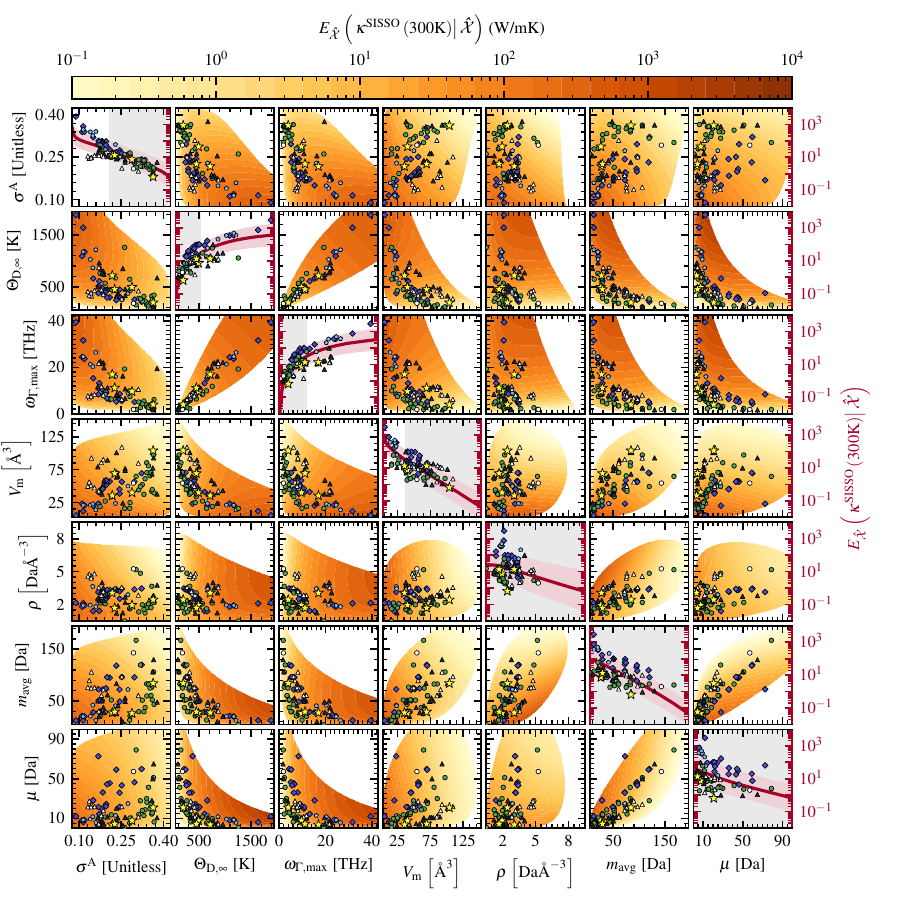}
    \caption{The expected value of $\kappaRTp$, $E_{\hat{\mathcal{X}}}\left(\left.\kappaRTp\right| \hat{\mathcal{X}} \right)$, where $\hat{\mathcal{X}}$ is  defined by the variables on the $x-$ and $y-$axes. All one-dimensional maps are shown along the diagonal, where the $y$-axis represents the expected value of the models (the red axis on the right). $E_{\hat{\mathcal{X}}}\left(\left.\kappaRTp\right| \hat{\mathcal{X}} \right)$ is calculated by sampling over the multivariate distributions used for the sensitivity analysis, and binning the input data until there are at least 10,000 samples in each bin. The red line in the diagonal plots corresponds to $E_{\hat{\mathcal{X}}}\left(\left.\kappaRTp\right| \hat{\mathcal{X}} \right)$ and the pink shaded region is one standard deviation on either side of the line. The gray shaded regions represent where a thermal conductivity of 10 Wm$^{-1}$K$^{-1}$ or lower is within one standard deviation of the expected value. On all maps all materials in the training set are displayed. The green circles correspond to rock-salts, the blue diamonds are zincblende, the light blue pentagons are wurtzites, and black triangles are all other materials. All points with a $\kappaRT$ less than one standard deviation below the expected value based on $\sigmaA$ are highlighted in white. The points for the diagonal plots correspond to the actual values of $\kappaRT$ for each material. Additionally we include four new materials outside of the training set (yellow stars) whose thermal conductivities we calculate using \textit{ab initio} molecular dynamics.}
    \label{fig:si_all_maps}
\end{figure}

\FloatBarrier

\subsection{Comparison of Sensitivity Analysis Results with and without Modelling Input Dependence}

\begin{figure*}
    \centering
    \includegraphics{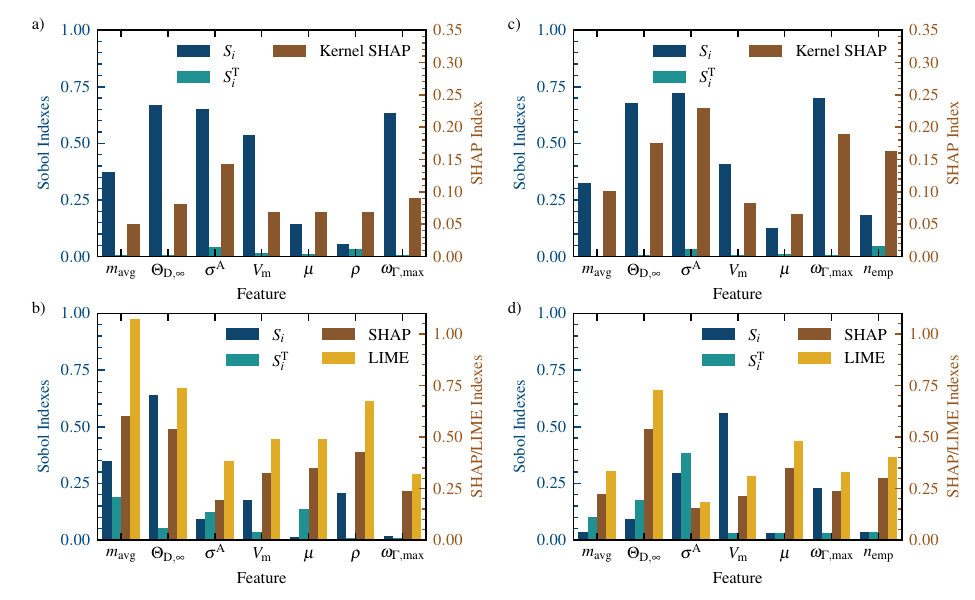}
    \caption{Demonstration of the impact of simplification of the discovered expression. a-b) The value of the sensitivity index for each primary feature in Equation~1 ($x-$axis). c-d)  The value of the sensitivity index for each primary feature where $\frac{\rho V_\mathrm{m}}{m_\mathrm{avg}}$ in Equation~1 is replaced by the number of atoms in the empirical formula, $n_\mathrm{emp}$. $S_i$ (first bar, dark blue), $S_i^\mathrm{T}$ (second bar, light blue), mean absolute SHAP index (third bar, brown), and LIME index (fourth bar, yellow) for each feature in the model by treating the inputs as a and c) dependent feature and b and d) independent features.}
    \label{fig:si_sensitivity_analysis}
\end{figure*}

Supplementary Figure~\ref{fig:si_sensitivity_analysis} confirms that the high values for $m_\mathrm{avg}$ seen in Figure~\ref{fig:sensitivity_analysis}b are likely an artifact of sampling over physically inaccessible regions of the input space, namely when $\frac{\rho V_\mathrm{m}}{m_\mathrm{avg}} \ne n_\mathrm{emp}$, where $n_\mathrm{emp}$ is the number of atoms in the empirical formula.
If we replace $\frac{\rho V_\mathrm{m}}{m_\mathrm{avg}}$ with $n_\mathrm{emp}$ in Equation~\ref{eqn:model} and recalculate the various metrics in Supplementary Figure~\ref{fig:si_sensitivity_analysis}d, we see a significant drop off in the importance of $m_\mathrm{avg}$ for all metrics.
While for $m_\mathrm{avg}$ it was possible to partially decouple the inputs by simplifying the formula, this is not always the case.
For example, $\omega_\mathrm{\Gamma,max}$ and $\Theta_\mathrm{D,\infty}$ are highly correlated to each other, but there is no clear simplification that removes this correlation.
In fact, the ability to do this at all represents a key advantage of symbolic regression techniques: the ability to directly interrogate the found expression.
Without analyzing Equation~\ref{eqn:model} and finding the $n_\mathrm{emp}$ simplification, the discrepancy between the various importance metrics for $m_\mathrm{avg}$ would have remained unresolved.
This analysis demonstrates the need to include correlative effects for this problem.
\FloatBarrier

\subsection{Training New Models Including Features from the Slack Model}
\label{sec:si_slack_training}
\begin{figure}[htb!]
    \centering
    \includegraphics{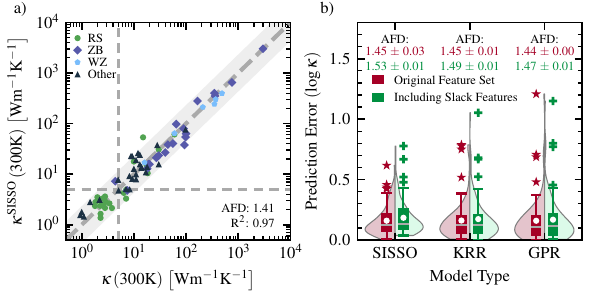}
    \caption{a) A comparison of the predicted $\kappaRTp$ against the measured $\kappaRT$ for the model trained against the updated dataset including the features from the Slack model. Gray shaded region corresponds to 95\% confidence interval. b) Violin plots of the mean prediction error of all samples for the SISSO, KRR, and GPR models using the main feature set (red, left) and the one including the features from the Slack model (green, right). Gray lines are the median and white circles are the mean of the distributions, the boxes are the quartiles, and the whiskers are the minimum and 95\% absolute error metrics. The red hexagons and green pluses are all points outside of the whiskers. }
    \label{fig:si_slack_training}
\end{figure}
As an additional check on the performance of the generated models we retrain the SISSO, KRR, and GPR models including three additional features from the Slack model, $\Theta_\mathrm{a}$, $\gamma$, and $V_\mathrm{a}$.
Under the assumption that heat is transported only by acoustic modes and that only Umklapp processes contribute to phonon scattering, the Slack model approximates $\kappaL$  as
\begin{equation}
    \kappaL = A \frac{{m_\mathrm{avg} V_\mathrm{a}^{1/3} n_\mathrm{at}^{1/3} \Theta_\mathrm{a}^3}}{T \gamma^2} \;.
    \label{eq:slack}
\end{equation}
Here, $m_\mathrm{avg}$ is the average mass of the atoms in the primitive cell; $V_\mathrm{a}$ is the atomic volume; $\Theta_a = {\Theta_\mathrm{D}}/{\sqrt[3]{n_\mathrm{at}}}$ is the Debye temperature of the acoustic modes; $\Theta_\mathrm{D}$ is the Debye temperature; $T$ is the temperature; $\gamma$ is the high-temperature, thermodynamic Gr\"uneisen parameter; $n_\mathrm{at}$ are the number of atoms in the primitive cell; and $A$ is a fitting constant approximated as~\cite{Slack1979,Julian1965}:
\begin{equation}
    A = \frac{2.43\times10^{-6} }{1-0.514/\gamma+0.228/\gamma^2} \mathrm{\frac{Kg}{Da} \frac{m}{\text{\AA}} \frac{1}{K^3 s^3}}.
\end{equation}
All models are generated using the same procedure as outlined in Section~\ref{sec:method}.
Supplementary Figure~\ref{fig:si_slack_training} illustrates the performance of the new models for both the training and prediction error.
While the training error for the individual one and two-dimensional models are better than those in the main text, the additional features lead to an increased prediction error for all models.
Because of this, the optimal model found by SISSO is one-dimensional
\begin{equation}
    \log \left(\kappaRTpss\right) = a_0 + a_1\left(\frac{\Vm}{V_\mathrm{a} \ln\left(\mu + 22.72\, \mathrm{Da}\right)} + \left(\sigmaA + 0.889\right)^3 - \ln\left(\omgG + 7.283 \,\mathrm{THz}\right) \right),
    \label{eqn:model_ss}
\end{equation}
where $a_0=0.4228$ and $a_1=-1.164$ .
This model mirrors the $d_2$ term in $\kappaRTp$, with the ratio $\frac{\rho}{m_\mathrm{avg}}$ being replaced by $\frac{1}{V_\mathrm{a}}$ with a slightly different dependence on $\sigmaA$ and $\omgG$.
This similarity between these models, as well as the increased prediction error shown in Figure~\ref{fig:si_slack_training}b indicate that including these new features does not produce the optimal models.

\begin{table}
\centering
\caption{Sensitivity Analysis results for the selected model}
\label{tab:si_SA_ss}
\begin{tabular}{AAAAAA}
\toprule
  & $V_\mathrm{a}$ & $\bm{\Vm}$ & $\mu$ & $\bm{\omgG}$ & $\bm{\sigmaA}$\\
\midrule
$S_i$     & 0.45 & \textbf{0.47} & 0.10 & \textbf{0.66} & \textbf{0.80} \\
$S_i^\mathrm{T}$ & 0.03 & \textbf{0.10} & 0.00 & \textbf{0.06} & \textbf{0.10} \\
KerSHAP   & 0.08 & \textbf{0.23} & 0.04 & \textbf{0.24} & \textbf{0.26} \\
\botrule
\end{tabular}
\end{table}

\begin{table}[htb!]
\centering
\caption{Summary of the univariate marginal distributions used to generate the synthetic data for the SISSO models using the Slack features. }
\label{tab:si_marginal_distribution_slack}
\begin{tabular}{ADD}
\toprule
  & Type & Parameters\\
\midrule
$V_\mathrm{a}$  & Gamma      & $\theta=12.6303$, $k=1.7522$   \\
$\Vm$           & Weibull    & $\lambda=58.4857$, $k=2.2902$  \\
$\mu$           & Gamma      & $\theta=4.872$, $k=4.4075$     \\
$\omgG$         & Log-Normal & $\mu = 2.0993$, $\sigma=0.608$ \\
$\sigmaA$       & Uniform    & $a=0.075$, $b=0.425$           \\
\botrule
\end{tabular}
\end{table}

\begin{table}[htb!]
\centering
\caption{The Pearson correlation parameters used for the Gaussian Copula for the SISSO models using the Slack features}
\label{tab:si_copula_slack}
\begin{tabular}{AAAAAA}
\toprule
  & $V_\mathrm{a}$ & $\mathbf{\Vm}$ & $\mu$ & $\bm{\omgG}$ & $\bm{\sigmaA}$\\
\midrule
$V_\mathrm{a}$ &  1.0    &  0.7703 &  0.7936 & -0.8538 &  0.6172 \\
$\Vm$          &  0.7703 &  1.0    &  0.4442 & -0.5880 &  0.4423 \\
$\mu$          &  0.7936 &  0.4442 &  1.0    & -0.7133 &  0.3019 \\
$\omgG$        & -0.8538 & -0.5880 & -0.7133 &  1.0    & -0.7490 \\
$\sigmaA$      &  0.6172 &  0.4423 &  0.3019 & -0.7490 &  1.0    \\
\botrule
\end{tabular}
\end{table}
As can be seen in Supplementary Table~\ref{tab:si_SA_ss}, performing sensitivity analysis on this new model gives similar results as what we saw in the main text with $\sigmaA$, $\Vm$, and $\omgG$ being the most important inputs.
For this case $\omgG$ is selected instead of $\ThetaD$, but as can be seen in Supplementary Figure~\ref{fig:si_all_maps} there are only slight differences in the information contained by these features.
Additionally, the high correlation between $V_\mathrm{a}$ and the other selected features shown in Supplementary Table~\ref{tab:si_copula_slack} is likely inflating its importance.

\begin{figure}[tbp!]
    \centering
    \includegraphics{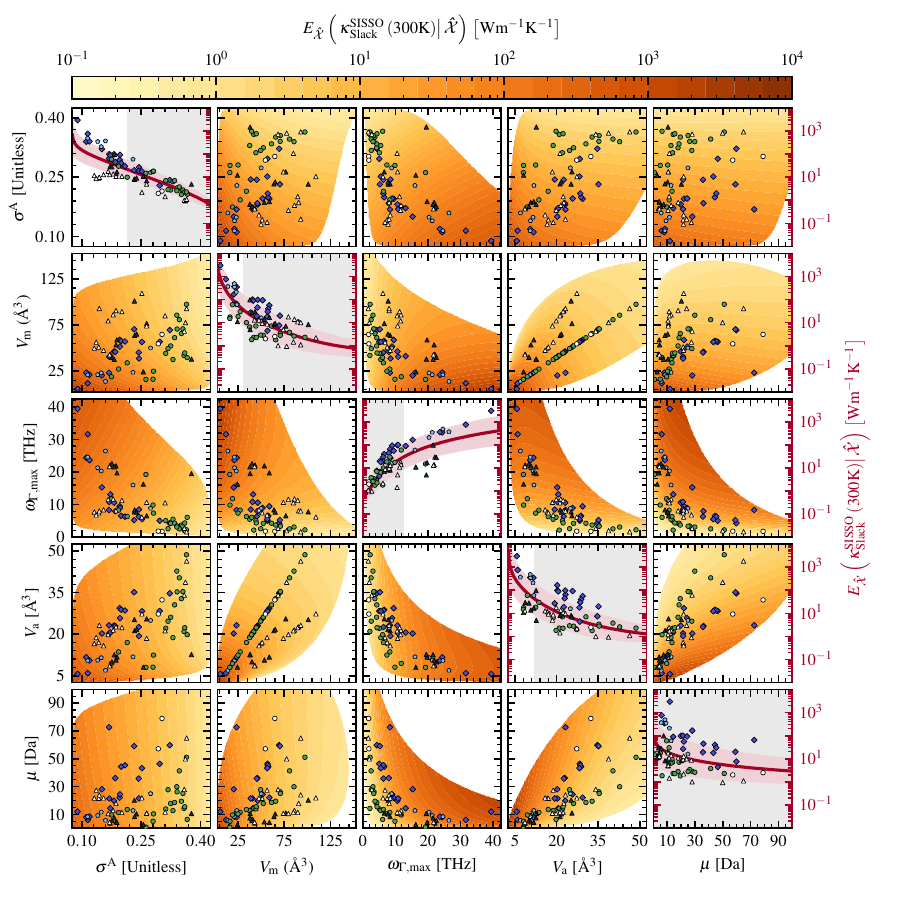}
    \caption{The expected value of $\kappaRTpss$, $E_{\hat{\mathcal{X}}}\left(\left.\kappaRTpss\right| \hat{\mathcal{X}} \right)$, where $\hat{\mathcal{X}}$ is  defined by the variables on the $x$ and $y$-axis. All one-dimensional maps are shown along the diagonal, where the $y$-axis represents the expected value of the models (the red axis on the right). $E_{\hat{\mathcal{X}}}\left(\left.\kappaRTpss\right| \hat{\mathcal{X}} \right)$ is calculated by sampling over the multivariate distributions used for the sensitivity analysis, and binning the input data until there are at least 10,000 samples in each bin. The red line in the diagonal plots corresponds to $E_{\hat{\mathcal{X}}}\left(\left.\kappaRTpss\right| \hat{\mathcal{X}} \right)$ and the pink shaded region is one standard deviation on either side of the line. The gray shaded regions represent where a thermal conductivity of 10 Wm$^{-1}$K$^{-1}$ or lower is within one standard deviation of the expected value. On all maps all materials in the training set are displayed. The green circles correspond to rock-salts, the blue diamonds are zincblende, the light blue pentagons are wurtzites, and black triangles are all other materials. All points with a $\kappaRT$ less than one standard deviation below the expected value based on $\sigmaA$ are highlighted in white. The points for the diagonal plots correspond to the actual values of $\kappaRT$ for each material.}
    \label{fig:si_all_maps_ss}
\end{figure}

The maps of this new model over all primary features are very similar to the ones generated for the models in the main text, as can be seen in Supplementary Figure~\ref{fig:si_all_maps_ss}.
The largest difference between the two sets of maps is the leveling off of $\kappaRTpss$ at $\sim$1 Wm$^{-1}$K$^{-1}$ for larger volumes and the flatter curve for $\mu$, which is attributed to losing the information contained in the $d_1$ term of $\kappaRTp$.
Additionally, there is a lower uncertainty for the expected value of $\kappaRTpss$ with respect to $\sigmaA$, particularly in the low thermal conductivity regime, which is likely from the simpler expression found in Equation~\ref{eqn:model_ss}.
Interestingly, the dependence of the model on $V_\mathrm{a}$, is the opposite of what one would expect upon inspecting the Slack model which suggests that $\kappaL$ increases with increasing $V_\mathrm{a}$.
However, there is a strong inverse correlation between $V_\mathrm{a}$ and $\Theta_\mathrm{a}$, which in turn inverts the relationship between $\kappaL$ and $V_\mathrm{a}$, highlighting the need to include
correlative effects when studying these systems.
Finally, when comparing the conditions for finding new thermal insulators proposed by these models we see only a slight deviation from the ones used in the main text, as shown in the gray shaded regions in Supplementary Figure~\ref{fig:si_all_maps_ss}.
\pagebreak

\bibliography{kappa_sig}
\bibliographystyle{naturemag}

\end{document}